\documentclass[10pt,twocolumn,aps,prd,superscriptaddress,nofootinbib,fleqn,showkeys,showpacs]{revtex4-2}
\usepackage{graphicx}
\usepackage{dcolumn}
\usepackage{bm}
\usepackage{hyperref}
\usepackage{newunicodechar}
\usepackage{amsmath,amssymb}
\usepackage{comment} 
\allowdisplaybreaks
\newunicodechar{−}{\ensuremath{-}}

\usepackage[dvipsnames]{xcolor}
\definecolor{darkblue}{rgb}{0.0, 0.0, 0.55}
\definecolor{darkred}{rgb}{0.55, 0.0, 0.0}
\usepackage{cleveref}
\hypersetup{
    colorlinks=true,
    linkcolor=darkblue,
    citecolor=darkblue,
    urlcolor=darkblue}

\newcommand{\XiExt}{\Xi}

\makeatletter
\let\@bibdataout@init\relax
\AtBeginDocument{%
  \let\write@bibliographystyle\relax
  \let\auto@bib\@empty
  \let\auto@bib@innerbib\@empty
  \let\bibliography\@gobble
  \let\bibliographystyle\@gobble
}
\makeatother

\begin{document}

%\title{Dynamical DE from extrinsic curvature beyond constant-$w$ cosmologies }

%\title{Time-dependent effective DE from the extrinsic curvature of embedded spacetime}

\title{Extrinsic curvature as a source of time-dependent effective dark energy}

\author{Abra\~{a}o J. S. Capistrano}
\email{capistrano@ufpr.br}
\affiliation{Departamento de Engenharia e Exatas, Universidade Federal do Paran\'{a}, Rua Pioneiro 2153, Palotina, 85950-000, Paran\'{a}/PR, Brazil}
\affiliation{Applied Physics Graduate Program, Federal University of Latin American Integration (UNILA), Avenida Tarqu\'{i}nio Joslin dos Santos 1000, Polo Universit\'{a}rio, Foz do Igua\c{c}u, 85867-670, Paran\'{a}/PR, Brazil}

\author{Emanuelly Silva}
\email{emanuelly.santos@ufrgs.br}
\affiliation{Instituto de F\'{i}sica, Universidade Federal do Rio Grande do Sul, 91501-970, Porto Alegre, RS, Brazil}

\author{Rafael C. Nunes}
\email{rafadcnunes@gmail.com}
\affiliation{Instituto de F\'{i}sica, Universidade Federal do Rio Grande do Sul, 91501-970, Porto Alegre, RS, Brazil}
\affiliation{Divis\~{a}o de Astrof\'{i}sica, Instituto Nacional de Pesquisas Espaciais, Avenida dos Astronautas 1758, S\~{a}o Jos\'{e} dos Campos, 12227-010, S\~{a}o Paulo, Brazil}

\author{Orlando Luongo}
\email{orlando.luongo@unicam.it}
\affiliation{Department of Nanoscale Science and Engineering, University at Albany-SUNY, Albany, New York 12222, USA}
\affiliation{Universit\`{a} di Camerino, Divisione di Fisica, Via Madonna delle Carceri 9, 62032 Camerino, Italy}
\affiliation{INFN, Sezione di Perugia, 06123 Perugia, Italy}
\affiliation{INAF, Osservatorio Astronomico di Brera, Milano, Italy}
\affiliation{Al-Farabi Kazakh National University, Al-Farabi Avenue 71, 050040 Almaty, Kazakhstan}

\author{Suresh Kumar}
\email{suresh.kumar@plaksha.edu.in}
\affiliation{Plaksha University, Mohali, Punjab 140306, India}

\date{\today}

\begin{abstract}
We extend the extrinsic-curvature cosmological scenario based on Nash-Greene embeddings beyond the constant equation of state (EoS) approximation. {We replace} the vacuum Ricci-flat Einstein-Gupta closure {with} a sourced auxiliary relation, {so that} the FLRW background {can be expressed in terms of} a single bending function $\alpha(a)=d\ln b/d\ln a$, which maps exactly onto a dynamical effective EoS, $w(a)=[1-2\alpha(a)]/3$, without introducing a dark energy (DE) fluid or scalar field. At the linear level, a phenomenological scalar extrinsic response, with amplitude $\beta_0$, induces a scale-dependent modification of the metric response. We confront the model with Planck CMB, DESI DR2 BAO, and the PantheonPlus, Union3, and DES-Dovekie supernova samples, considering two cases for the effective EoS: a constant $w$ ($w$CDM) and the CPL parametrization ($w_0w_a$CDM) of the bending history. The constant-$w$ extrinsic sector is consistent with the data, but is not favored over $\Lambda$CDM. {For the CPL parametrization, the data favor} a phantom divide crossing at $z\simeq0.4$ - $0.5$, with a moderate, supernova-dependent preference over $\Lambda$CDM for Union3 and DES-Dovekie, {but no corresponding preference} for PantheonPlus or CMB+DESI alone. The scalar amplitude is bounded by $\beta_0\lesssim0.3$ at 95\% CL, {without evidence for a nonzero amplitude}, mainly because the modification acts near the horizon scale.
\end{abstract}



\maketitle

\section{Introduction}\label{sec:introduction}

{The unresolved unification problem in fundamental physics motivates the investigation of geometric extensions of Einstein gravity.} Despite the {success of general relativity (GR) and the Standard Model of particle physics}, a complete theoretical framework in which gravitation and quantum gauge interactions arise from a common structure {remains} unavailable. Historically, this problem stimulated higher-dimensional constructions, beginning with Kaluza-Klein theories and later passing through supergravity, string-inspired models and brane-world scenarios. In such approaches, the extra dimensions are regarded as possible geometric carriers of gravitational dynamics. The {Arkani-Hamed-Dimopoulos-Dvali model}~\cite{ArkaniHamed1998}, the Randall-Sundrum model I (RS-I)~\cite{Randall1999a}, the Randall-Sundrum model II (RS-II)~\cite{Randall1999b} and the {Dvali-Gabadadze-Porrati model}~\cite{Dvali2000} are among the most influential realizations of this idea, since they connect the hierarchy problem with possible departures from four-dimensional gravity at short or cosmological scales. However, several brane-world models depend on rigid embeddings~\cite{Randall1999a,Randall1999b,Dvali2000}, mirror symmetries~\cite{Randall1999a,Randall1999b}, junction conditions~\cite{Lanczos1924,Israel1966,Battye2001,Davis2003} or fixed bulk ans\"atze~\cite{ArkaniHamed1998,Randall1999a,Randall1999b,Dvali2000}. {Subsequent studies showed that junction conditions are not necessary in a general brane-world framework}~\cite{Maia1984,Maia1989,Mukohyama2000,Deruelle2001,Gergely2003,Barcelo2000,Brax2004,Yamauchi2007,Jalalzadeh2009}. {Following this approach, we consider smooth deformations of the embedded spacetime along orthogonal directions, treating the extrinsic curvature as an independent geometric quantity instead of identifying it algebraically with the confined energy-momentum tensor, as commonly done in conventional brane-world models}~\cite{Randall1999a,Randall1999b}. We then adopt the Nash-Greene embedding framework~\cite{Nash1956,Greene1970} {to describe these deformations}. {Our model thus follows a dynamical embedding approach, grounded in the Nash–Greene framework, in which the extrinsic curvature is treated as a dynamical geometric quantity rather than being restricted through conventional brane-world junction conditions.}

Recent observational developments have also sharpened the {DE} problem. The {increasing precision of baryon acoustic oscillation (BAO) measurements from the Dark Energy Spectroscopic Instrument (DESI)}~\cite{DESI:2025zgx,DESI:2025fii}, together with {cosmic microwave background (CMB) measurements}~\cite{Giare:2024oil,AtacamaCosmologyTelescope:2025nti,Giare:2025pzu} and Type Ia supernova compilations~\cite{NatureDESI2025,Camilleri:2026rer}, has {renewed interest in the possibility of departures from cosmological constant behavior in the late-time expansion}. Full-shape clustering analyses further strengthen this program by jointly probing cosmic expansion and structure growth~\cite{DESI2024FS,Silva:2025twg,Chudaykin:2025lww,Chudaykin:2026nls}. At the same time, large-scale-structure measurements from the Kilo Degree Survey (KiDS), the Dark Energy Survey (DES)~\cite{Wright:2025xka,DES:2026mkc} and the Hyper Suprime-Cam (HSC)~\cite{HSCY3,LSSTDarkEnergyScience:2026ovp,DES:2026jmi} have played a central role in the discussion of the {amplitude of matter clustering and the associated possible observational tension}. Although these observations do not, by themselves, provide conclusive evidence against $\Lambda$CDM, they motivate extensions of the standard framework in which modifications to the background expansion and the growth of cosmic structures arise in a correlated manner, as naturally occurs in modified gravity models~\cite{Du:2026cly,Paliathanasis:2026vhi,Specogna:2024euz,Hartmann:2026vak,Montani:2025nmz,Lu:2025sjg,Efstratiou:2025iqi,Zheng:2025vyv,Akarsu:2026anp,SanchezLopez:2025uzw,Scherer:2025esj,DeFelice:2026led,Cadoni:2026cyh,Wolf:2025acj,Adam:2025kve,Mishra:2025goj,Mazumdar:2025smh,Hussain:2025vbo,Hogas:2025ahb,Ye:2025pem,Akarsu:2024qsi,Richarte:2026cza,Khoury:2026svx,Hashim:2026yoy,Reboucas:2026ljj,Valletta:2025bgu,Wang:2025znm,Liu:2025myr,Paliathanasis:2025hjw,Akarsu:2026lva,Nojiri:2025uew,Nojiri:2025low,Capistrano:2024kuc,Wen:2026xlj,Andrade:2023pws,Kumar:2022nvf,Shah:2025vnt,Ishak:2024jhs,Toda:2024fgv}.

To {address these questions, we consider the observed four-dimensional spacetime as an embedded manifold whose extrinsic curvature contributes dynamically to the gravitational field equations}. In the Nash {framework, the extrinsic curvature generates orthogonal deformations of the metric, promoting the embedding variables to dynamical quantities}. In this sense, the extrinsic geometry can act as an effective DE sector without introducing an \emph{ad hoc} scalar degree of freedom, in the spirit of the embedded-cosmology program developed in Refs.~\cite{MAIA20029,GDE,gde2,Capistrano2020,Capistrano2021,Capistrano2026}.

In the embedded FLRW setting, the background contribution of the extrinsic curvature is {described by a bending function}. {In the vacuum Einstein-Gupta construction, the logarithmic rate of variation of the bending function with respect to the cosmological scale factor is constant.} This leads to a power-law extrinsic density and therefore to a constant effective EoS. This {constant EoS realization allowed the first assessment of the model's observational viability and the derivation of} the modified perturbation equations and the effective Newtonian coupling. However, current DE analyses often {consider time-dependent EoSs, motivating a geometrically consistent extension of the Nash framework in which this evolution follows from the bending dynamics}. {To investigate this possibility, we replace the vacuum Ricci-flat Einstein-Gupta closure with a sourced auxiliary relation, where the source parametrizes the unresolved geometric response of the embedding without introducing an additional physical matter component. Under this assumption, we derive the corresponding background evolution and scalar perturbative response, and confront the resulting constant and dynamical bending histories with cosmological observations.}

{The paper is organized as follows. In Sec.~\ref{sec:nash_baseline}, we summarize the Nash-Greene embedding framework and the constant EoS background. In Sec.~\ref{sec:bending_to_wz}, we introduce the sourced auxiliary Gupta relation and derive the dynamical bending history. In Sec.~\ref{sec:perturbation_system}, we construct the scalar perturbation system, while in Sec.~\ref{sec:cpl_ba}, we express its gravitational response through the modified growth and lensing functions and introduce the Chevallier-Polarski-Linder (CPL) parametrization}~\cite{CPL1,CPL2}. {In Sec.~\ref{sec:obs_numerical}, we describe the observational datasets and the statistical methodology. In Sec.~\ref{sec:results}, we present the constraints on the constant and dynamical scenarios and discuss their cosmological implications. Finally, in Sec.~\ref{conclusioni}, we summarize the main results and possible perspectives.}

\section{Nash-Greene embeddings in FLRW Cosmology}\label{sec:nash_baseline}

We summarize the main findings following Refs.~\cite{MAIA20029,GDE,gde2,Gupta,Capistrano2020,Capistrano2021,Capistrano2026}. Starting from a four-dimensional FLRW space-time embedded in a five-dimensional bulk $^5{\mathcal R}_{ABCD}$, the four-dimensional induced field equations on the space-time are
\begin{subequations}\label{eq:induced_background}
\begin{align}
\bar{G}_{\mu\nu}-\bar{Q}_{\mu\nu}
   &=-8\pi G\bar{T}_{\mu\nu},
   \label{eq:noperttensoreq_v5}\\
\bar{k}_{\mu[\nu;\rho]}&=0.
   \label{eq:nopervecteq_v5}
\end{align}
\end{subequations}
together with the confinement condition for the bulk matter source,
\begin{subequations}\label{eq:confinement_v5}
\begin{align}
8\pi G \bar{T}_{\mu\nu}
  =&G_*\mathcal{Z}^A_{,\mu}
    \mathcal{Z}^B_{,\nu}T^*_{AB},&\\
\mathcal{Z}^A_{,\mu}\eta^B T^*_{AB}&=0,&\\
\eta^A\eta^B T^*_{AB}&=0,&
\end{align}
\end{subequations}
where $G$ and $G_\ast$ are the Newtonian and the higher-dimensional Newtonian gravitational constant, respectively. These geometries, i.e., the bulk and embedded space, must be compatible with the integrability conditions. From the Nash smooth perturbations~\cite{Nash1956}, as a solution for the embedding problem, one gets the known Gauss and Codazzi equations~\cite{Greene1970} given by
\begin{eqnarray}
\nonumber&&^5{\cal R}_{ABCD}\mathcal{Z}^A_{,\alpha}\mathcal{Z}^B_{,\beta}\mathcal{Z}^C_{,\gamma}
\mathcal{Z}^D_{,\delta}= \bar{R}_{\alpha\beta\gamma\delta}\;+\\
&&\hspace{4cm}+\;(\bar{k}_{\alpha\gamma}\;\bar{k}_{\beta\delta}\!-\!\;\bar{k}_{\alpha\delta}\;\bar{k}_{\beta\gamma})\!\;,\label{eq:G1}\\
&&^5{\cal R}_{ABCD}\mathcal{Z}^A_{,\alpha} \mathcal{Z}^B_{,\beta}\mathcal{Z}^C_{,\gamma}\eta^D=\;\bar{k}_{\alpha[\beta;\gamma]} \;, \label{eq:C1}
\end{eqnarray}
where $^5{\cal R}_{ABCD}$ is the five-dimensional Riemann tensor, and $\bar{R}_{\alpha\beta\gamma\delta}$ is the four-dimensional background Riemann tensor. The square brackets denote antisymmetrization over the enclosed indices and semicolon symbol denotes covariant derivative. Then, the embedding is now well established with the perturbed embedding coordinate that is denoted by $\mathcal{Z}^{A}_{,\mu}$ and is defined through the Nash-Greene deformation process as
\begin{equation}
\mathcal{Z}^{A}_{,\mu}=\mathcal{X}^{A}_{,\mu}+\delta y\,\eta^{A}_{,\mu}\;,
\end{equation}
where $\mathcal{X}^{A}_{,\mu}$ represents the tangent basis of the non-perturbed embedded space-time, $\eta^{A}$ is the unit vector field orthogonal to the embedded hypersurface, and $\delta y$ denotes an infinitesimal deformation parameter along the extra-dimensional direction. This setting follows from the Lie transport of the Gaussian frame along the orthogonal flow generated by $\eta^{A}$, characterizing the smooth local deformations of the embedded geometry. Aligned with Nash embedding theorem~\cite{Nash1956}, the perturbations here are performed exclusively along the normal direction of the hypersurface which guarantees that the deformation process remains gauge-independent with respect to intrinsic coordinate transformations on the embedded space-time. Moreover, the normal vector $\eta^{A}$ remains invariant under the perturbation flow as $\eta^{A} \rightarrow \eta^{A}$, since the Lie bracket $[\eta,\eta]^{A}$ identically vanishes. As a consequence, the orthogonality and normalization conditions of the Gaussian frame are preserved during the deformation process, maintaining the consistency of the embedding structure under Nash-Greene fluctuations~\cite{MAIA20029,GDE,gde2,Capistrano2026}.

The deformation tensor $Q_{\mu\nu}$ in Eq.~\eqref{eq:noperttensoreq_v5} is given by
\begin{equation}
\begin{split}
\bar{Q}_{\mu\nu}={}
\bar{k}^{\rho}_{\mu}\bar{k}_{\rho\nu}
-\bar{k}_{\mu\nu}h-\frac{1}{2}\left(K^2-h^2\right)\bar{g}_{\mu\nu} &,
\end{split}
\label{eq:qmunu_v5}
\end{equation}
and carries the extrinsic-curvature contribution terms, where $h=\bar g^{\mu\nu}\bar k_{\mu\nu}$ is the mean Gaussian curvature and $K^2=\bar k^{\mu\nu}\bar k_{\mu\nu}$ is the Gaussian curvature. We point out that the tensor $\bar{Q}_{\mu\nu}$ is conserved in the sense that $\bar{Q}_{\mu\nu;\mu}=0$. Throughout this work, capital Latin indices run from 1 to 5 and denote quantities in the embedding space. Lowercase Latin indices refer to the one extra dimension considered. Greek indices span from 1 to 4 and are associated with the embedded four-dimensional space-time. A bar over a symbol denotes a background (unperturbed) quantity. Moreover, we adopt the Landau-Lifshitz metric signature convention, $ (-,+,+,+) $.

As shown in \cite{GDE,gde2}, for the background cosmology, we use the spatially flat FLRW metric,
\begin{equation}
  ds^2= -dt^2 + a^2\left(dr^2+r^2d\theta^2 + r^2\sin^2\theta d\phi^2 \right)\;.
\end{equation}
As solution of Eqs.\eqref{eq:noperttensoreq_v5} and \eqref{eq:nopervecteq_v5}~\cite{GDE}, 
the extrinsic geometry is written in terms of the bending function $b(t)$ and the ratio $B/H$, with $H=\dot a/a$ and $B=\dot b/b$, and one has the components
\begin{subequations}\label{eq:BB_v5}
\begin{align}
\bar{k}_{ij}&=\frac{b}{a^2}\bar{g}_{ij},
& i,j&=1,2,3,\\
\bar{k}_{44}&=-\frac{b}{a^{2}}\left(\frac{B}{H}-1\right),\\
K^{2}&=\frac{b^2}{a^4}
\left(\frac{B^2}{H^2}-2\frac BH+4\right),
& h&=\frac{b}{a^2}\left(\frac BH+2\right),\\
\bar{Q}_{ij}&=\frac{b^{2}}{a^{4}}
\left(2\frac{B}{H}-1\right)\bar{g}_{ij},
&\bar{Q}_{44}&=-\frac{3b^{2}}{a^{4}},\\
\bar{Q}&=\frac{6b^{2}}{a^{4}}\frac{B}{H} .
\end{align}
\end{subequations}
Moreover, to determine the dynamics of $b(t)$, one then normalizes the extrinsic curvature via
\begin{equation}
f_{\mu\nu} = \frac{2}{K}k_{\mu\nu},
\label{eq:fmunu_v5}
\end{equation}
that introduces the associated $f$-connection and $f$-curvature tensor as shown in \cite{gde2}. In the vacuum Einstein-Gupta realization \cite{Gupta}, the corresponding Ricci tensor satisfies
\begin{equation}
{\cal  F}_{\mu\nu}=0 \;,
\label{eq:guptaflat_v5}
\end{equation}
which leads, for the FLRW case, to the constant ratio solution
\begin{equation}
\begin{aligned}
\frac{B}{H}&=\alpha_0,\qquad
b(t)=b_0 a^{\alpha_0}(t).
\end{aligned}
\label{eq:constantalpha}
\end{equation}
where $b_0$ and $\alpha_0$ are integration constants. Then, the background Friedmann equation becomes
\begin{equation}
E^2(z)\equiv \left(\frac{H}{H_0}\right)^2
=\Omega_{m0}(1+z)^3+\Omega_{{\rm ext},0}(1+z)^{4-2\alpha_0},
\label{eq:constantFriedmann}
\end{equation}
with $\Omega_{{\rm ext},0}=1-\Omega_{m0}$ for a flat universe. In analogy with an effective cosmological fluid $8\pi G\,T^{\rm ext}_{\mu\nu}\equiv \bar{Q}_{\mu\nu}$ it leads to the expressions
\begin{equation}\label{eq:ext_rho_p}
\begin{aligned}
\bar{\rho}_{\rm ext}(a)
 &=\bar{\rho}_{{\rm ext},0}a^{2\alpha_0-4},\\
\bar{p}_{\rm ext}(a)
 &=\frac{1-2\alpha_0}{3}
   \,\bar{\rho}_{{\rm ext},0}a^{2\alpha_0-4}.
\end{aligned}
\end{equation}
The former quantities $\bar{\rho}_{\rm ext}(a)$ and $\bar{p}_{\rm ext}(a)$ denote the effective energy density and pressure associated with the extrinsic geometric sector induced by the embedding of the four-dimensional space-time into the higher-dimensional bulk. The parameter $\bar{\rho}_{{\rm ext},0}$ corresponds to the present-day value of the effective extrinsic energy density at $a=1$, while $\alpha_0$ is a dimensionless parameter controlling the dynamical evolution of the extrinsic contribution along the cosmological expansion. 

From Eq.~\eqref{eq:ext_rho_p}, the effective EoS parameter of the extrinsic sector is obtained as
\begin{equation}\label{eq:constEoS}
w_{\rm ext}
=w=
\frac{\bar{p}_{\rm ext}}{\bar{\rho}_{\rm ext}}
=
\frac{1-2\alpha_0}{3},
\end{equation}
showing that the parameter $\alpha_0$ determines the deviation of the extrinsic geometric contribution from the standard matter- or $\Lambda$-like behaviours. This constant-$w$ limit is the baseline scenario used in the previous analyses of the model Refs.~\cite{Capistrano2020,Capistrano2021,Capistrano2026}. In the following section, we show how it generalizes when the bending ratio is allowed to vary with time.

\section{From the bending function $b(t)$ to a dynamical EoS $w(z)$}\label{sec:bending_to_wz}

A varying EoS follows once the constant vacuum solution in Eq.~(\ref{eq:constantalpha}) is relaxed. Hereafter, we adopt the conservative interpretation that the effective Gupta sector departs from the strict vacuum Ricci-flat regime. This departure is shown by the response of the extrinsic-curvature dynamics to the intrinsic FLRW curvature and by effective geometric contributions induced by the embedding dynamics. Such interpretation is also consistent with the general idea that geometric sectors beyond the standard Einstein term can induce effective cosmological sources, as occurs in averaging approaches, brane-world projections, and modified-gravity descriptions of DE \cite{Buchert2000,Maartens2010,Nojiri2011}.

At the background level, this generalization leads to relaxing the
Ricci-flat Gupta condition of Eq.~\eqref{eq:guptaflat_v5}. In the original
Gupta construction, the auxiliary tensor $f_{\mu\nu}$ is introduced only
as a tensorial structure associated with the extrinsic curvature
$k_{\mu\nu}$ as shown in Eq.~\eqref{eq:fmunu_v5}. We stress that it should not be interpreted as a second physical metric:
the spacetime geometry probed by observers is still determined by
$g_{\mu\nu}$. The role of $f_{\mu\nu}$ is to provide a convenient
spin-2 representation in which a dynamical equation for the extrinsic
curvature can be written. The physical consequences of this auxiliary
dynamics are then transferred to the four-dimensional cosmology through
the evolution of $k_{\mu\nu}$, and subsequently through the geometric
tensor $Q_{\mu\nu}$~\cite{GDE,gde2}.

As previously noticed, the vacuum choice in Eq.~\eqref{eq:guptaflat_v5} corresponds to the simplest Ricci-flat closure of the Gupta-sector dynamics. However, the source of the auxiliary Gupta equation is not fixed \textit{a priori}, since the dynamics of spacetime deformations is not known independently of cosmological observations. Hence, we replace the
vacuum closure by the effective sourced relation
\begin{equation}\label{eq:gupta1}
{\cal F}_{\mu\nu}={\cal S}_{\mu\nu},
\end{equation}
where ${\cal S}_{\mu\nu}$ is not an additional matter stress-energy
tensor in the physical spacetime. Rather, it parametrizes the unresolved
geometric response of the extrinsic-curvature sector, including possible
intrinsic-curvature backreaction, nonlinear deformation effects, and
boundary information inherited from the higher-dimensional embedding.
In this sense, ${\cal S}_{\mu\nu}$ is a source for the auxiliary
spin-2 deformation equation, not a direct source for the physical
Einstein equations.

Assuming homogeneity and isotropy, the full tensorial structure of
${\cal S}_{\mu\nu}$ does not need to be specified independently. Then the
FLRW symmetry reduction closes the background dynamics in terms of a
single deformation function,
\begin{equation}\label{eq:alphadef}
\alpha(a)\equiv \frac{d\ln b}{d\ln a},
\end{equation}
which controls the evolution of the extrinsic curvature sector.  Hence,
$\alpha(a)$ measures departures from the vacuum Gupta regime and
determines the amplification of the effective extrinsic geometric
contribution throughout the cosmological expansion. Thus, Eq.~(\ref{eq:alphadef}) integrates immediately to
\begin{equation}
b(a)=b_0 \exp\!\left[\int_{1}^{a}\alpha(\tilde a)\,d\ln \tilde a\right].
\label{eq:bofa}
\end{equation}
It is convenient to define the purely geometric amplification factor
\begin{equation}
\XiExt(a)\equiv \exp\!\left[2\int_{1}^{a}\alpha(\tilde a)\,d\ln \tilde a\right]
=\left[\frac{b(a)}{b_0}\right]^2,
\label{eq:Xidef}
\end{equation}
which directly quantifies the cumulative growth of the extrinsic sector relative to its present-day amplitude $b_0$. Consequently, the effective extrinsic energy density evolves according to
\begin{equation}
\bar\rho_{\rm ext}(a)=\bar\rho_{{\rm ext},0}\,a^{-4}\,\XiExt(a),
\label{eq:rhovar}
\end{equation}
which reduces to $\bar\rho_{{\rm ext},0}a^{2\alpha_0-4}$ when $\alpha(a)=\alpha_0$.  It
shows that the standard radiation-like contribution $a^{-4}$ is geometrically modified by the amplification factor $\XiExt(a)$. Therefore, we obtain the corresponding effective EoS in a form
\begin{equation}
w_{\rm ext}(a)
=
\frac{1-2\alpha(a)}{3},
\end{equation}
establishing a direct correspondence between the deformation dynamics and the effective fluid-like behaviour of the extrinsic sector. It is worth noting that the constant-$\alpha_0$ configuration is recovered as the particular limit
\begin{equation}\label{eq:Xi}
\XiExt(a)=a^{2\alpha_0},
\end{equation}
for which we obtian the relations in Eq.~\eqref{eq:constantalpha}. In the vacuum Gupta limit of Eq.~\eqref{eq:guptaflat_v5}, we have
\begin{equation}
{\cal S}_{\mu\nu}\rightarrow 0,
\qquad
\alpha(a)\rightarrow \alpha_0,
\end{equation}
then the deformation sector reduces to the original Ricci-flat configuration with constant extrinsic EoS parameter.

We are able now to update the background density and pressure for a dynamical EoS. Since the background extrinsic energy-momentum tensor is conserved, one has
\begin{equation}
\dot{\bar\rho}_{\rm ext}+3H\left(\bar\rho_{\rm ext}+\bar p_{\rm ext}\right)=0.
\label{eq:extcons}
\end{equation}
Using Eqs.~(\ref{eq:BB_v5}) and (\ref{eq:Xidef}), the exact density reduces to Eq. (\ref{eq:rhovar}). Then, the corresponding pressure follows from the continuity equation,
\begin{equation}
w_{\rm ext}(a)\equiv w(a)\equiv \frac{\bar p_{\rm ext}}{\bar\rho_{\rm ext}}
=-1-\frac{1}{3}\frac{d\ln \bar\rho_{\rm ext}}{d\ln a}.
\end{equation}
Combining with Eq.~(\ref{eq:rhovar}), it gives the exact mapping
\begin{equation}
w(a)= -1+\frac{1}{3}\bigl[4-2\alpha(a)\bigr],
\label{eq:wfromalpha}
\end{equation}
or equivalently
\begin{equation}
\alpha(a)=\frac{1-3w(a)}{2}.
\label{eq:alphafromw}
\end{equation}
This is the full functional generalization of Eq.~(\ref{eq:constEoS}). Once either $\alpha(a)$ or $w(a)$ is specified, the other is fixed algebraically.

Using Eqs.(\ref{eq:rhovar}) - (\ref{eq:alphafromw}), the Hubble rate closes exactly as
\begin{equation}
E^2(a)=\Omega_{m0}a^{-3}+\Omega_{{\rm ext},0}a^{-4}\XiExt(a),
\label{eq:E2Xi}
\end{equation}
or, equivalently, in the more familiar fluid form
\begin{equation}
\begin{split}
E^2(z)={}&\Omega_{m0}(1+z)^3\\
&+\Omega_{{\rm ext},0}
\exp\!\left[3\int_{0}^{z}
\frac{1+w(z')}{1+z'}dz'\right].
\end{split}
\label{eq:E2w}
\end{equation}
Then equations~(\ref{eq:bofa}) - (\ref{eq:E2w}) constitute the exact background closure of the model.

We also obtain complementary set of background relations directly associated with the generalized extrinsic sector. First, taking the logarithmic derivative of Eq.~\eqref{eq:E2Xi} with the prime notation $'=d/da$, and using the conservation equations for both matter and extrinsic components, one obtains
\begin{equation}
\frac{H'}{H}=
-\frac{3\Omega_m(a)+\Omega_{\rm ext}(a)\left[4-2\alpha(a)\right]}{2a},
\label{eq:Hprime}
\end{equation}
where the fractional energy densities are defined by
\begin{equation}
\Omega_i(a)\equiv
\frac{\rho_i(a)}
{\rho_m(a)+\rho_{\rm ext}(a)}.
\end{equation}
This expression explicitly shows how departures from the standard FLRW evolution are governed by the deformation function $\alpha(a)$ through the effective extrinsic contribution. Moreover, since the effective extrinsic sector behaves as a generalized dynamical fluid whose EoS parameter evolves according to Eq.~\eqref{eq:wfromalpha}, the associated adiabatic sound speed is written as
\begin{equation}
c^2_{a,{\rm ext}}(a)\equiv
\frac{\dot{\bar p}_{\rm ext}}
{\dot{\bar\rho}_{\rm ext}}
=
w(a)
-\frac{a\,w'(a)}
{3[1+w(a)]},
\label{eq:ca2}
\end{equation}
whenever $1+w\neq 0$. This quantity plays an important role in determining the stability and propagation properties of linear perturbations associated with the effective extrinsic fluid description, but  not introduced as an independent perturbative closure across the phantom divide. Here we adopt the geometrized density $(8\pi G/3)\rho_{{\rm ext}}$, so that
$\bar\rho_{{\rm ext},0}=H_0^2\Omega_{{\rm ext},0}$.

\section{The scalar perturbation system}
\label{sec:perturbation_system}

We now construct the scalar perturbation system associated with the Nash-Greene deformation sector. Unlike the conventional description of DE in terms of an effective fluid, the present framework attributes the additional cosmological contribution to the extrinsic geometry of the embedded spacetime. Accordingly, the corresponding perturbative source is constructed from the first variation of the geometrical deformation tensor $Q_{\mu\nu}$, without introducing an independent dark energy fluid closure.

\subsection{Linearized Gupta--Codazzi system}

We work in the conformal Newtonian gauge, in which the perturbed line element reads
\begin{equation}
\begin{split}
ds^2=a^2(\tau)
\big[-(1+2\Psi)d\tau^2
+(1-2\Phi)\delta_{ij}dx^i dx^j\big],
\end{split}
\label{eq:scalarpertmetric2_v6}
\end{equation}
where $\Psi$ is the Newtonian potential and $\Phi$ represents the scalar perturbation of the spatial curvature. Throughout this section, an overdot denotes differentiation with respect to conformal time $\tau$, while a prime denotes differentiation with respect to the scale factor $a$. The conformal Hubble rate is defined as $\mathcal H\equiv\dot a/a=aH$.

Since both the metric and the extrinsic curvature are perturbed, it is convenient to linearize the induced field equation in Eq.~\eqref{eq:noperttensoreq_v5} directly in mixed-index form,
\begin{equation}
\delta G^\mu{}_\nu
-
\delta Q^{(\beta,s)\mu}{}_\nu
=
-8\pi G\,\delta T^\mu{}_\nu.
\label{eq:fieldpert_linear}
\end{equation}
Hereafter, an overbar denotes a background quantity, the superscript $(s)$ identifies the scalar perturbation sector, and the superscript $(1,s)$ denotes a unit-normalized scalar deformation response. The superscript $(\beta,s)$ identifies the corresponding physical response with dimensionless amplitude $\beta_0$. These superscripts are mode labels and should not be confused with tensor indices or perturbative orders.

The construction of $\delta Q^{(\beta,s)\mu}{}_\nu$ requires specifying how the extrinsic geometry responds to scalar perturbations of the embedded spacetime. To this end, we extend the sourced Gupta description introduced in the previous section, Eq.~\eqref{eq:gupta1}, to the linear perturbation regime. The background configuration is governed by the sourced Gupta equation
\begin{equation}
\overline{\mathcal F}_{\mu\nu}
[\bar f(\bar k)]
=
\overline{\mathcal S}_{\mu\nu},
\label{eq:sourced_gupta_background}
\end{equation}
where $\overline{\mathcal S}_{\mu\nu}$ represents unresolved higher-dimensional, boundary, and nonlinear-deformation contributions. Its role is to support the homogeneous extrinsic configuration and, consequently, the cosmological bending history characterized by $\alpha(a)$.

The first-order scalar variation of Eq.~\eqref{eq:sourced_gupta_background} gives
\begin{equation}
\delta\mathcal F^{(s)}_{\mu\nu}
[\delta k^{(s)}]
=
\delta\mathcal S^{(s)}_{\mu\nu}.
\label{eq:linear_sourced_gupta}
\end{equation}
In the absence of an explicit dynamical description of the fluctuations of the effective higher-dimensional source, we adopt the minimal scalar-source closure
\begin{equation}
\delta\mathcal S^{(s)}_{\mu\nu}=0.
\label{eq:minimal_scalar_source_closure}
\end{equation}

This assumption does not eliminate the background source $\overline{\mathcal S}_{\mu\nu}$, which continues to determine the homogeneous bending history. Rather, it restricts the perturbative analysis to the response governed by the homogeneous linearized Gupta equation, without introducing additional source degrees of freedom associated with the unresolved embedding dynamics. This constitutes a minimal modeling assumption rather than a direct consequence of the embedding theorem.

Under this prescription, the scalar extrinsic-curvature perturbation is subject to the homogeneous linearized Gupta equation and the linearized Codazzi constraint,
\begin{equation}
\delta\mathcal F^{(s)}_{\mu\nu}
[\delta k^{(s)}]=0,
\qquad
\delta^{(s)}
\!\left(k_{\mu[\nu;\rho]}\right)=0.
\label{eq:homogeneous_gupta_codazzi}
\end{equation}
The latter expression denotes the first variation of the complete Codazzi relation, including the perturbation of the connection associated with the induced metric.

The system in Eq.~\eqref{eq:homogeneous_gupta_codazzi} provides the dynamical and geometrical conditions that admissible scalar extrinsic deformations must satisfy. Its complete solution space may contain several scalar modes with distinct tensorial structures and initial conditions. For the phenomenological analysis developed here, however, we restrict attention to a single scalar response branch.

We now introduce a dimensionless amplitude $\beta_0$ for the selected extrinsic response. To account for the metric-induced contribution to the mixed shape operator, we normalize the departure $\Delta k_{\mu\nu}$ from the reference deformation defined as
\begin{equation}
\Delta k^{(\beta,s)}_{\mu\nu}=\beta_0\,\Delta k^{(1,s)}_{\mu\nu}.
\label{eq:gupta_mode_beta}
\end{equation}
Moreover, to establish the connection between this extrinsic response and the perturbed gravitational field equations, we recall that $Q_{\mu\nu}$ is a geometrical functional of both the induced metric and the extrinsic curvature. Its complete mixed-index first variation is consequently defined by
\begin{equation}
\delta Q^{(\beta,s)\mu}{}_\nu
\equiv
\left.
\frac{d}{d\epsilon}
Q^\mu{}_\nu
\big[g(\epsilon),k(\epsilon;\beta_0)\big]
\right|_{\epsilon=0},
\label{eq:deltaQ_beta_definition}
\end{equation}
where $\epsilon$ is the perturbative expansion parameter.

\subsection{Geometrical deformation and mixed shape operator}

Since the mixed-index variation depends on both $\delta g_{\mu\nu}$ and $\delta k_{\mu\nu}$, the reference deformation must be separated before normalizing the response. From the FLRW extrinsic-curvature components in Eq.~\eqref{eq:BB_v5}, the quadratic invariant entering the deformation tensor in Eq.~\eqref{eq:qmunu_v5} takes the form
\begin{equation}
\bar K^2-\bar h^2
=
-6\alpha(a)\frac{b^2(a)}{a^4},
\label{eq:K2h2_sectionIV}
\end{equation}
or, equivalently,
\begin{equation}
\bar K^2-\bar h^2
=
-6b_0^2a^{-4}\XiExt(a)\alpha(a).
\label{eq:K2h2_Xi_sectionIV}
\end{equation}

It is useful to introduce the geometrized background amplitude such as
\begin{equation}
\mathcal U(a)
\equiv
b_0^2a^{-4}\XiExt(a).
\label{eq:U_definition}
\end{equation}
In terms of this quantity, the nonvanishing background mixed components of the deformation tensor are
\begin{equation}
\bar Q^4{}_4
=
3\mathcal U(a),
\qquad
\bar Q^i{}_j
=
[2\alpha(a)-1]\mathcal U(a)\delta^i{}_j,
\label{eq:Q_background_mixed}
\end{equation}
where the index $4$ denotes the conformal-time component. Hence, $\mathcal U(a)$ determines the overall geometrical amplitude, whereas $\alpha(a)$ controls the relative contributions of the temporal and spatial components.

The relation between metric and extrinsic-curvature perturbations can be examined starting from the normal deformation of the embedded spacetime. In the Nash deformation framework \cite{Nash1956}, the metric variation associated with an infinitesimal displacement $\delta y$ along the normal direction is
\begin{equation}
\delta g_{\mu\nu}
=
-2\delta y\,\bar k_{\mu\nu}.
\label{eq:nash_delta_g_sectionIV}
\end{equation}
For the particular normal deformation considered here, the corresponding extrinsic-curvature variation is written as
\begin{equation}
\begin{split}
\delta k_{\mu\nu}
%&=%-2\delta y\,\bar g^{\rho\sigma}\bar k_{\mu\rho}\bar k_{\nu\sigma}
&=
\bar g^{\rho\sigma}
\bar k_{\mu\rho}\delta g_{\nu\sigma}.
\end{split}
\label{eq:nash_delta_k_sectionIV}
\end{equation}
These expressions provide a reference deformation against which a more general scalar response of the extrinsic geometry can be characterized.

To make this distinction explicit, we introduce the mixed shape operator
\begin{equation}
A^\mu{}_\nu
\equiv
g^{\mu\rho}k_{\rho\nu},
\end{equation}
in terms of which the extrinsic-curvature invariants are given by
\begin{equation}
h=\operatorname{tr}(A),
\qquad
K^2=\operatorname{tr}(A^2).
\end{equation}
Their first variations are
\begin{align}
\delta A^\mu{}_\nu
&=
\bar g^{\mu\rho}\delta k_{\rho\nu}
+
\delta g^{\mu\rho}\bar k_{\rho\nu},
\\
\delta h=&\operatorname{tr}(\delta A),\;\delta K^2=2\operatorname{tr}(\bar A\delta A).
\end{align}
Consequently, the mixed-index deformation tensor in Eq.~\eqref{eq:qmunu_v5} can be expressed as
\begin{equation}
Q^\mu{}_\nu
=
(A^2)^\mu{}_\nu
-
hA^\mu{}_\nu
-
\frac{1}{2}(K^2-h^2)\delta^\mu{}_\nu,
\end{equation}
and its complete first variation reads
\begin{align}
\delta Q^\mu{}_\nu
={}&
(\delta A\bar A+\bar A\delta A)^\mu{}_\nu
-
\bar h\,\delta A^\mu{}_\nu
\nonumber\\
&-
\bar A^\mu{}_\nu\delta h
-
\frac{1}{2}
(\delta K^2-2\bar h\delta h)
\delta^\mu{}_\nu.
\label{eq:general_deltaQ_mixed}
\end{align}
Equation~\eqref{eq:general_deltaQ_mixed} shows that the mixed-index deformation response is entirely determined by the perturbation of the shape operator. In particular, using the normal variations in Eqs.~\eqref{eq:nash_delta_g_sectionIV} and \eqref{eq:nash_delta_k_sectionIV}, together with the inverse-metric identity we have
\begin{equation}
\delta g^{\mu\nu}
=
-\bar g^{\mu\rho}
\bar g^{\nu\sigma}
\delta g_{\rho\sigma}.
\end{equation}
And one obtains
\begin{equation}
\bar g^{\mu\rho}\delta k_{\rho\nu}
=
-\delta g^{\mu\rho}\bar k_{\rho\nu}.
\end{equation}
The two contributions to $\delta A^\mu{}_\nu$ therefore cancel, yielding
\begin{equation}
\delta A^\mu{}_\nu=0, \quad \delta h=0, \quad \delta K^2=0.
\end{equation}
It follows immediately from Eq.~\eqref{eq:general_deltaQ_mixed} that
\begin{equation}
\left.\delta Q^\mu{}_\nu
\right|_{\rm normal}
=0.
\label{eq:pure_normal_Q_response}
\end{equation}
This result does not imply, however, that the covariant deformation tensor remains unchanged. Indeed, varying the relation $Q_{\mu\nu}=g_{\mu\rho}Q^\rho{}_\nu$ gives
\begin{equation}
\delta Q_{\mu\nu}
=
\bar g_{\mu\rho}\delta Q^\rho{}_\nu
+
\bar Q^\rho{}_\nu\delta g_{\mu\rho}.
\label{eq:Q_mixed_covariant_relation}
\end{equation}
Thus, even when the mixed shape operator is unchanged, the covariant tensor retains a contribution associated with the variation of the metric used to lower its index.

The distinction between these two contributions is particularly relevant for cosmological perturbations. Since the gravitational field equations are written in mixed-index form, a nonvanishing scalar deformation source requires a variation of the shape operator beyond the particular normal displacement described above. Such an additional response is precisely what the extrinsic Gupta dynamics is intended to accommodate.

The general structure of this additional deformation can be characterized by introducing
\begin{equation}
\Delta k_{\mu\nu}
\equiv
\delta k_{\mu\nu}
-
\bar g^{\rho\sigma}
\bar k_{\mu\rho}\delta g_{\nu\sigma}.
\label{eq:additional_extrinsic_response}
\end{equation}
For the diagonal scalar metric perturbations considered here, the variation of the shape operator then reduces to
\begin{equation}
\delta A^\mu{}_\nu=\bar g^{\mu\rho}\Delta k_{\rho\nu}.
\label{eq:deltaA_additional_response}
\end{equation}
Accordingly, $\Delta k_{\mu\nu}$ measures the departure from the reference normal deformation and determines the nontrivial mixed-index response through Eq.~\eqref{eq:general_deltaQ_mixed}.

\subsection{Scalar response and physical normalization}

A complete determination of $\Delta k_{\mu\nu}$ would require solving the coupled scalar Gupta-Codazzi system. In the present work, rather than introducing additional independent scalar response functions, we adopt a restricted phenomenological description of this sector. Specifically, we consider a local scalar response that preserves the diagonal and spatially isotropic tensorial structure of the conformal Newtonian metric perturbation.

Within this description, the intrinsic deformation of the mixed tensor and the metric-induced contribution in Eq.~\eqref{eq:Q_mixed_covariant_relation} are required to combine into a covariant response proportional to the metric perturbation. For the unit-response branch, we therefore prescribe
\begin{equation}
\delta Q^{(1,s)}_{\mu\nu}
=
\mathcal C_Q(a)\delta g_{\mu\nu}.
\label{eq:deltaQ_covariant_result}
\end{equation}

The background-dependent coefficient $\mathcal C_Q(a)$ must have the same geometrical dimensions as the deformation tensor. A natural background quantity with these dimensions is the quadratic extrinsic-curvature invariant $\bar K^2-\bar h^2$, which already appears explicitly in the definition of $Q_{\mu\nu}$. We consequently retain a response proportional to this invariant and adopt the normalization
\begin{equation}
\begin{split}
\mathcal C_Q(a)
&\equiv
-\frac{3}{2}
\left(\bar K^2-\bar h^2\right)\\
&=
9\alpha(a)\mathcal U(a).
\end{split}
\label{eq:CQ_definition}
\end{equation}
We point out that Equations~\eqref{eq:deltaQ_covariant_result} and \eqref{eq:CQ_definition} specify a phenomenological reference response without additional time-dependent functions.

Having specified the covariant response, we now return to the mixed-index formulation required by the perturbed field equations. Varying the inverse metric in Eq.~\eqref{eq:Q_mixed_covariant_relation} and substituting the unit-response prescription in Eq.~\eqref{eq:deltaQ_covariant_result}, we obtain
\begin{equation}
\begin{split}
\delta Q^{(1,s)\mu}{}_\nu
={}&
\mathcal C_Q(a)
\bar g^{\mu\rho}\delta g_{\rho\nu}-\bar g^{\mu\rho}\delta g_{\rho\sigma}\bar Q^\sigma{}_\nu.
\end{split}
\label{eq:unit_response_mixed_general}
\end{equation}
This equation establishes the direct connection between the selected covariant scalar response and the mixed-index geometrical source entering the gravitational field equations.

Although the covariant response is characterized by a single coefficient $\mathcal C_Q(a)$, the temporal and spatial components of the background deformation tensor are generally different. As a result, the variation of the inverse metric generates distinct temporal and spatial response coefficients.

We now need to express our result explicitly in terms of the perturbed Newtonian potentials $\Psi$, and $\Phi$. From the perturbed line element in Eq.~\eqref{eq:scalarpertmetric2_v6}, the nonvanishing covariant metric perturbations are
\begin{subequations}
\begin{align}
\delta g_{44}&=-2a^2\Psi,\\
\delta g_{4i}&=0,\\
\delta g_{ij}&=-2a^2\Phi\,\delta_{ij}.
\end{align}
\label{eq:metric_pert_components}
\end{subequations}
Using these expressions together with the background components in Eq.~\eqref{eq:Q_background_mixed}, the mixed-index unit-response components follow directly from Eq.~\eqref{eq:unit_response_mixed_general},
\begin{align}
\delta Q^{(1,s)4}{}_4
&=
2\left[
\mathcal C_Q(a)-3\mathcal U(a)
\right]\Psi,
\label{eq:deltaQ_unit_44}
\\
\delta Q^{(1,s)4}{}_i
&=0,
\label{eq:deltaQ_unit_4i}
\\
\delta Q^{(1,s)i}{}_j
&=
-2\left\{
\mathcal C_Q(a)
-
[2\alpha(a)-1]\mathcal U(a)
\right\}
\Phi\,\delta^i{}_j.
\label{eq:deltaQ_unit_ij}
\end{align}

These expressions motivate the introduction of two background-dependent deformation-response functions, namely,
\begin{subequations}
\begin{align}
\mathcal M_Q(a)
&\equiv
\mathcal C_Q(a)-3\mathcal U(a)
\nonumber\\
&=
3\mathcal U(a)[3\alpha(a)-1],
\label{eq:MQ_definition}
\\
\mathcal P_Q(a)
&\equiv
\mathcal C_Q(a)
-
[2\alpha(a)-1]\mathcal U(a)
\nonumber\\
&=
[7\alpha(a)+1]\mathcal U(a).
\label{eq:PQ_definition}
\end{align}
\end{subequations}
The functions $\mathcal M_Q(a)$ and $\mathcal P_Q(a)$ characterize, respectively, the temporal and isotropic spatial responses of the selected deformation sector. They are not independent perturbative degrees of freedom or additional closure relations. Rather, both functions are completely specified by the homogeneous extrinsic geometry once the scalar response prescription has been adopted.

From Eqs.~\eqref{eq:gupta_mode_beta}, \eqref{eq:deltaA_additional_response} and \eqref{eq:general_deltaQ_mixed}, the normalization of $\Delta k_{\mu\nu}$ implies $\delta A^{(\beta,s)\mu}{}_{\nu}=\beta_0\delta A^{(1,s)\mu}{}_{\nu}$ and hence
\begin{equation}
\delta Q^{(\beta,s)\mu}{}_\nu
=
\beta_0\,
\delta Q^{(1,s)\mu}{}_\nu.
\label{eq:deltaQ_mode_linearity}
\end{equation}
Then the normalization leaves $\bar Q^\mu{}_{\nu}$ unchanged and gives $\delta Q^{(\beta,s)\mu}{}_{\nu}=0$ for $\beta_0=0$. Therefore, the resulting deformation tensor can be written in the compact form in terms of the Newtonian potentials as
\begin{subequations}
\begin{align}
\delta Q^{(\beta,s)4}{}_4
&=
2\beta_0\mathcal M_Q(a)\Psi,
\label{eq:deltaQ44_component}
\\
\delta Q^{(\beta,s)4}{}_i
&=
0,
\label{eq:deltaQ4i_component}
\\
\delta Q^{(\beta,s)i}{}_j
&=
-2\beta_0\mathcal P_Q(a)
\Phi\,\delta^i{}_j.
\label{eq:deltaQij_component}
\end{align}
\label{eq:deltaQ_components}
\end{subequations}
Within the selected scalar response prescription, the extrinsic deformation sector contributes directly to the temporal and isotropic spatial-trace gravitational equations. It is worth noticing that the vanishing off-diagonal temporal-spatial component and the absence of a traceless spatial contribution imply that this branch introduces neither an independent scalar momentum source nor intrinsic scalar anisotropic stress.

\subsection{Bianchi consistency}

At the background level, the Codazzi equation in
Eq.~\eqref{eq:C1}, together with its contracted form in
Eq.~\eqref{eq:nopervecteq_v5}, ensures the conservation
of the deformation tensor,
$\bar\nabla_\mu\bar Q^\mu{}_\nu=0$.
Using Eq.~\eqref{eq:Q_background_mixed}, the temporal
component of this conservation identity reduces to
$\dot{\mathcal U}+2\mathcal H(2-\alpha)\mathcal U=0$,
which is identically satisfied by the definition of
$\mathcal U$ in Eq.~\eqref{eq:U_definition}, while the
spatial components vanish by homogeneity. This result is independent of $\beta_0$. At the perturbative level, for a Gauss--Codazzi-compatible family, we have
\begin{equation}
\nabla^{(\epsilon)}_\mu
Q^{\mu}{}_{\nu}
\big[g(\epsilon),k(\epsilon;\beta_0)\big]=0.
\label{eq:family_Q_conservation}
\end{equation}
Differentiating Eq.~\eqref{eq:family_Q_conservation} at $\epsilon=0$ gives the complete linearized identity
\begin{equation}
\begin{split}
&\delta\!\left(\nabla_\mu Q^{\mu}{}_{\nu}\right)_{(\beta,s)}=0.
\end{split}
\label{eq:linear_bianchi_explicit}
\end{equation}

%my note:The complete first variation includes the connection terms, namely $\bar\nabla_\mu\delta Q^\mu{}_{\nu}+\delta\Gamma^\mu_{\mu\lambda}\bar Q^\lambda{}_{\nu}-\delta\Gamma^\lambda_{\mu\nu}\bar Q^\mu{}_{\lambda}$. A particular exact consistency limit of Eq.~\eqref{eq:deltaQ_components} is $\alpha=2$ at all times and $\beta_0=0$. In this limit $\mathcal U$ is constant, $\bar Q^\mu{}_{\nu}=3\mathcal U\delta^\mu{}_{\nu}$, and the connection terms cancel identically while the mixed-index perturbation vanishes. Thus the background contribution survives and the full linearized conservation identity holds. Away from this limit, the diagonal reference response is not generically conserved,i.e, for nonzero Fourier modes its spatial conservation condition is $\beta_0(7\alpha+1)\Phi_k=(\alpha-2)\Psi_k$, with a further temporal condition. The present parametrization is therefore tested as an effective metric-response prescription, not as a conserved tensor solution throughout the sampled parameter space.

The sourced Gupta equation must satisfy the corresponding
auxiliary contracted identity at both the background and
linear perturbation levels. Although the minimal source
prescription sets $\delta\mathcal S^{(s)}_{\mu\nu}=0$,
the first variation of this identity still involves
contributions from the perturbed auxiliary inverse metric
and connection, whose compatibility with the selected
scalar branch must be independently established. On the other hand, on the embedded spacetime, the confined matter sector
satisfies the standard covariant conservation law,
\begin{equation}
\nabla_\mu T^\mu{}_{\nu}=0,
\label{eq:matter_conservation_covariant}
\end{equation}
which, together with the contracted Bianchi identity,
requires the conservation of the geometrical deformation
tensor. At the linear perturbation level, this requirement
provides the corresponding consistency condition for
the scalar response entering Eq.~\eqref{eq:fieldpert_linear}.

Another important aspect to verify is the perturbation of the Codazzi equation in Eq.~\eqref{eq:nopervecteq_v5}. Consistently with the Bianchi treatment above, we retain the first variation of the complete covariant relation of Eq.~\eqref{eq:nopervecteq_v5},
\begin{equation}
\delta\!\left(k_{\mu\nu;\rho}-k_{\mu\rho;\nu}\right)=0.
\label{eq:codazzi_pert_sectionIV}
\end{equation}
This constraint includes the perturbed connection and must be satisfied together with the auxiliary Gupta equation in any geometrical realization of the selected response.

For comparison with fluid notation, one may define
\begin{equation}
8\pi G\delta T^{(Q)\mu}{}_\nu
\equiv-\delta Q^{(\beta,s)\mu}{}_\nu.
\label{eq:TQ_def_sectionIV}
\end{equation}
And we obtain the corresponding effective density and pressure perturbations as
\begin{subequations}
\label{eq:Q_effective_sources}
\begin{gather}
\delta\rho_Q=
\frac{\beta_0\mathcal M_Q(a)}{4\pi G}\Psi,
\label{eq:delta_rho_Q}\\[3pt]
\delta p_Q=
\frac{\beta_0\mathcal P_Q(a)}{4\pi G}\Phi,
\label{eq:delta_p_Q}\\[3pt]
(\bar\rho_Q+\bar p_Q)\theta_Q=0,\\
(\bar\rho_Q+\bar p_Q)\sigma_Q=0.
\end{gather}
\end{subequations}
At the phantom divide, we have an important aspect. Since $w(a)=-1$ implies $\alpha(a)=2$, so that $\mathcal M_Q(a)=\mathcal P_Q(a)=15\mathcal U(a)$. For a regular bending history these coefficients remain finite, without division by $1+w$. Accordingly, the retained algebraic metric response is regular at finite nonzero $k$ at the crossing for $\beta_0\geq0$ and positive $\mathcal U$. This removes the apparent singularity associated with the effective adiabatic sound speed, but does not by itself establish dynamical stability or conservation of the selected branch.

\subsection{Matter conservation and scalar field equations}

As usual, the ordinary matter sector obeys the standard conservation equations. Using the momentum-density variable $V=(1+w)\theta$, one has
\begin{subequations}
\label{eq:matter_conservation_sectionIV}
\begin{align}
\delta' &= 3(1+w)\Phi'
-\frac{V}{a^2H} -\frac{3}{a}
\left(\frac{\delta P}{\bar\rho}-w\delta\right),\\[3pt]
V' &= -\frac{1-3w}{a}V
+\frac{k^2}{a^2H}\frac{\delta P}{\bar\rho}+(1+w)\frac{k^2}{a^2H}(\Psi-\sigma).
\end{align}
\end{subequations}
For pressureless matter, we have
\begin{subequations}
\begin{align}
\delta'_m&=-\frac{\theta_m}{a^2H}+3\Phi',\\
\theta'_m&=-\frac{\theta_m}{a}+\frac{k^2}{a^2H}\Psi.
\end{align}
\label{eq:cdm_conservation_sectionIV}
\end{subequations}

In Fourier space, insertion of Eq.~\eqref{eq:deltaQ_components} into the field equations gives
\begin{equation}
k^2\Phi_k+3\mathcal H(\dot\Phi_k+\mathcal H\Psi_k)
=-4\pi Ga^2\delta\rho_k-\beta_0a^2\mathcal M_Q(a)\Psi_k,
\label{eq:einstein00_explicit_deltaQ}
\end{equation}
and the the momentum constraint
\begin{equation}
\dot\Phi_k+\mathcal H\Psi_k
=4\pi Ga^2(\bar\rho+\bar p)\frac{\theta_k}{k^2}.
\label{eq:momentum_constraint_deltaQ}
\end{equation}
In addition, the trace of the spatial components gives the dynamical
equation for the scalar metric potentials,
\begin{equation}
D_k+\frac{k^2}{3}(\Phi_k-\Psi_k)
=4\pi Ga^2\delta p_k+\beta_0a^2\mathcal P_Q(a)\Phi_k,
\label{eq:spatial_trace_deltaQ}
\end{equation}
where we denote
\begin{equation}
D_k\equiv\ddot\Phi_k+\mathcal H(2\dot\Phi_k+\dot\Psi_k)
+(\mathcal H^2+2\dot{\mathcal H})\Psi_k.
\label{eq:Dkdef}
\end{equation}
The traceless equation remains unchanged such as
\begin{equation}
k^2(\Phi_k-\Psi_k)=12\pi Ga^2(\bar\rho+\bar p)\sigma_k.
\label{eq:anisotropic_deltaQ}
\end{equation}
Hence $\Phi_k=\Psi_k$ when ordinary anisotropic stress is negligible. We finally obtain the main perturbed field Eqs.~\eqref{eq:einstein00_explicit_deltaQ}-\eqref{eq:anisotropic_deltaQ}. Next, we combine these constraints directly, without imposing either a quasi-static or a sub-horizon approximation.

\section{Nash gravity in the $\mu$-$\Sigma$ parametrization}\label{sec:cpl_ba}

The deformation history is reconstructed through the relations $
w(a)\leftrightarrow \alpha(a)\leftrightarrow \Xi(a)$, given by Eqs.\eqref{eq:Xidef}, \eqref{eq:rhovar} and \eqref{eq:alphafromw}. Moreover, the perturbative response appearing in the exact relativistic metric constraints is given by Eq.~\eqref{eq:MQ_definition}. For the sake of notation, it is now convenient to introduce the function
\begin{equation}
B(a)\equiv \beta_0a^2\mathcal M_Q(a).
\label{eq:B_nash_definition}
\end{equation}
The appearance of $B(a)$ is rooted in the scalar deformation response prescribed in Sec.~\ref{sec:perturbation_system}, motivated by the homogeneous Gupta-Codazzi system. The sourced Gupta equation and the linearized Codazzi equation provide the conditions for an admissible extrinsic response, while $\beta_0$ fixes the amplitude of the selected phenomenological branch. The resulting mixed-index perturbation of $Q^{\mu}{}_{\nu}$ enters the $00$ constraint algebraically, without an additional spatial Laplacian. It then appears beside the Einstein term $k^2\Psi_k$. This derivative mismatch gives rise to the scale dependence in $\mu$, while compatibility with the linearized conservation identity remains subject to the conditions discussed in the Bianchi consistency subsection.
Using the complete $00$, $0i$, and traceless spatial equations of Sec.~\ref{sec:perturbation_system}, one obtains the exact relativistic constraint
\begin{equation}
[k^2+B(a)]\Psi_k
=-4\pi Ga^2\left[\bar\rho\Delta+3(\bar\rho+\bar p)\sigma\right],
\label{eq:exact_psi_constraint_nash}
\end{equation}
where we denote
\begin{equation}
\bar\rho\Delta\equiv\delta\rho
+3\mathcal H(\bar\rho+\bar p)\frac{\theta}{k^2}.
\label{eq:total_comoving_density_secV}
\end{equation}
Therefore the exact relativistic modified-growth function is
\begin{equation}
\mu(a,k)=\frac{k^2}{k^2+B(a)}
=\left[1+\frac{\beta_0a^2\mathcal M_Q(a)}{k^2}\right]^{-1}.
\label{eq:mu_exact_nash}
\end{equation}
We point out that the scale dependence of $\mu(a,k)$ is not generated by a sub-horizon approximation. It follows directly from the operator structure of the full Fourier-space constraint. i.e., the Einstein tensor supplies the spatial Laplacian term $k^2\Psi_k$, whereas the algebraic mixed-index deformation response $\delta Q^{(\beta,s)\mu}{}_{\nu}$ supplies the time-dependent term $B(a)\Psi_k$. Solving the exact equation of Eq.~\eqref{eq:exact_psi_constraint_nash} necessarily produces the ratio $k^2/[k^2+B(a)]$. Thus, $k$ enters through the relativistic spatial-gradient operator, while $B(a)$ carries the geometric response of the extrinsic sector. The result is then an exact metric-response function, not a quasi-static effective Newton constant.

Another important point is that the Nash sector does not generate an intrinsic traceless scalar source. Hence the exact slip equation remains the same such as
\begin{equation}
k^2(\Phi_k-\Psi_k)=12\pi Ga^2(\bar\rho+\bar p)\sigma_k.
\label{eq:exact_slip_secV}
\end{equation}
Defining $\Pi\equiv(\bar\rho+\bar p)\sigma$, the exact Weyl-potential equation becomes as usual
\begin{equation}
k^2(\Phi_k+\Psi_k)
=-4\pi Ga^2\left[2\mu\bar\rho\Delta+(6\mu-3)\Pi\right].
\label{eq:exact_weyl_nash}
\end{equation}
If one formally writes this equation in the standard phenomenological form, we have
\begin{equation}
k^2(\Phi_k+\Psi_k)
=-4\pi Ga^2\Sigma_{\rm eff}
\left[2\bar\rho\Delta+3\Pi\right],
\end{equation}
then the closed relation
\begin{equation}
\Sigma_{\rm eff}(a,k)
=\mu(a,k)+
\frac{3[\mu(a,k)-1]\Pi}
{2\bar\rho\Delta+3\Pi}.
\label{eq:Sigma_exact_nash}
\end{equation}
Equation~\eqref{eq:Sigma_exact_nash} shows that, whenever ordinary photon or neutrino anisotropic stress is relevant, the exact lensing response is source dependent and cannot be represented by a function of $a$ and $k$ alone. In the late-time regime, where $\Pi\simeq0$, the exact system reduces to
\begin{equation}
\Sigma(a,k)=\mu(a,k),\qquad \Phi_k=\Psi_k.
\label{eq:mu_sigma_late_nash}
\end{equation}

For transparency in our numerical implementation, $b_0^2$, introduced in Eq.~\eqref{eq:constantalpha}, is absorbed into the present extrinsic-density normalization such as
\begin{equation}
b_0^2=H_0^2\Omega_{{\rm ext},0},
\end{equation}
where $H_0$ must be expressed in inverse-length units when it is combined with the CAMB wavenumber $k$. Thus we have
\begin{equation}
\mu(a,k)=
\left[
1+3\beta_0\Omega_{{\rm ext},0}
\left(\frac{H_0/c}{k}\right)^2
 a^{-2}\Xi(a)[3\alpha(a)-1]
\right]^{-1}.
\label{eq:mu_nash_numerical}
\end{equation}

\subsection{CPL parametrization}
\label{sec:CPL}
Having expressed the scalar sector through $\mu(a,k)$ and $\Sigma(a,k)$,
we now specify the bending history. Since the mapping
$w(a)\leftrightarrow\alpha(a)\leftrightarrow\Xi(a)$ is exact, any choice of $w(a)$ fixes
both the expansion history and the scalar response, with no additional
material DE component. As a first, controlled test of a time-dependent
bending history, we adopt the CPL parametrization
\cite{CPL1,CPL2},
\begin{equation}
  w(a)=w_0+w_a(1-a),
  \label{eq:cpl}
\end{equation}
which, in this framework, is the lowest-order smooth departure from the
constant-$\alpha$ solution around $a=1$, and allows direct comparison with
recent BAO, CMB and supernova analyses that test departures from a
cosmological constant \cite{NatureDESI2025}. Here, however, $w_0$ and
$w_a$ describe the evolution of the extrinsic bending sector, namely
$\alpha_0=(1-3w_0)/2$ and $d\alpha/d\ln a|_{a=1}=3w_a/2$, rather than those of an independently postulated
DE fluid.

Using the exact relation in Eq.~\eqref{eq:alphafromw}, one obtains
\begin{equation}
\alpha_{\rm CPL}(a)
=
\frac{1-3w_0-3w_a(1-a)}{2},
\label{eq:alpha_cpl_a}
\end{equation}
or, equivalently, in terms of redshift \textit{z}
\begin{equation}
\alpha_{\rm CPL}(z)
=
\frac{1-3w_0}{2}
-\frac{3w_a}{2}\frac{z}{1+z}.
\label{eq:alpha_cpl_z}
\end{equation}
The corresponding geometrical amplification factor follows directly
from Eq.~\eqref{eq:Xidef}, and we have
\begin{equation}
\XiExt_{\rm CPL}(a)
=
a^{1-3(w_0+w_a)}
\exp[3w_a(a-1)],
\label{eq:Xicpl_a}
\end{equation}
or, in terms of redshift \textit{z}
\begin{equation}
\XiExt_{\rm CPL}(z)
=
(1+z)^{3(w_0+w_a)-1}
\exp\left[-\frac{3w_az}{1+z}\right].
\label{eq:Xicpl_z}
\end{equation}
Consequently, the background expansion rate becomes
\begin{equation}
\begin{split}
E^2_{\rm CPL}(z)={}&
\Omega_{m0}(1+z)^3\\
&+\Omega_{{\rm ext},0}
(1+z)^{3(1+w_0+w_a)}
e^{-3w_az/(1+z)}.
\end{split}
\label{eq:Ecpl}
\end{equation}

The importance of this reconstruction is that the same CPL bending
history also fixes the scalar response of the extrinsic geometry.
Substituting $\alpha_{\rm CPL}(a)$ and
$\XiExt_{\rm CPL}(a)$ into the general expression for
$\mu(a,k)$ yields
\begin{equation}
\begin{aligned}
\mu_{\rm CPL}(a,k)
&=
\Bigg[
1+\frac{3}{2}\beta_0\Omega_{{\rm ext},0}
\left(\frac{H_0/c}{k}\right)^2
a^{-2}\XiExt_{\rm CPL}(a)
\\
&\;\;\qquad\times
\left(1-9w_{\rm ext}(a)\right)
\Bigg]^{-1}.
\end{aligned}
\label{eq:mu_cpl_exact}
\end{equation}
Thus, the same pair $(w_0,w_a)$ that controls the homogeneous
bending history also determines, together with $\beta_0$, the
time- and scale-dependent modification of the metric response. At
late times, when the anisotropic stress of the standard species is
negligible, one has
\begin{equation}
\Sigma_{\rm CPL}(a,k)=\mu_{\rm CPL}(a,k).
\label{eq:sigma_cpl_late}
\end{equation}
During radiation domination and around recombination, however, the
source-dependent Weyl-potential relation
Eq.~\eqref{eq:exact_weyl_nash} must be retained rather than imposing
a source-independent $\Sigma(a,k)$.

We point out that the CPL realization must also satisfy the physical conditions required
for a viable extrinsic DE sector. In particular, we require
$\Omega_{{\rm ext},0}>0$, $E^2(a)>0$, and accelerated expansion at
late times. Suppression of the extrinsic density relative to matter
at early times requires
\begin{equation}
w_0+w_a<0.
\label{eq:cpl_early_density_condition}
\end{equation}
For a nonzero constant $\beta_0$, recovery of the GR perturbative
limit $\mu(a,k)\rightarrow1$ at fixed $k$ as $a\rightarrow0$ further
requires
\begin{equation}
w_0+w_a<-\frac13.
\label{eq:cpl_early_gr_condition}
\end{equation}
Finally, regularity of the modified-gravity response requires
\begin{equation}
1+3\beta_0\Omega_{{\rm ext},0}
\left(\frac{H_0/c}{k}\right)^2
a^{-2}\XiExt_{\rm CPL}(a)
[3\alpha_{\rm CPL}(a)-1]>0
\label{eq:cpl_mu_regularity}
\end{equation}
over the complete $(a,k)$ domain explored by the Boltzmann evolution.

\section{Methodology}
\label{sec:obs_numerical}

This section describes the observational data and the statistical framework adopted to constrain the cosmological models considered in this study. We explore three distinct scenarios: i) the standard $\Lambda$CDM model, ii) Nash gravity + wCDM, and iii) Nash gravity + CPL \cite{CPL1,CPL2}. To test the theoretical frameworks, we implement them in the Boltzmann solver code \texttt{MGCAMB} \cite{Zhao:2008bn, Hojjati:2011ix, Zucca:2019xhg, Wang:2023tjj, Lewis:1999bs} and use the sampling tool \texttt{Cobaya} \cite{Torrado:2020dgo, 2019ascl.soft10019T} to perform Markov Chain Monte Carlo (MCMC) analyses, ensuring a Gelman-Rubin convergence criterion~\cite{Gelman_1992} of $R-1 \leq 0.02$ in all runs.

The priors assumed for each model are summarized in Table \ref{tab:priors}. We fix the neutrino mass at $\sum m_{\nu}=0.06\,{\rm eV}$ for all cases. Once the Markov chains were obtained, all post-processing was carried out using the \texttt{GetDist} package\footnote{Available at \url{https://github.com/cmbant/getdist}.}~\cite{Lewis:2019xzd} to extract numerical results.

\begin{table}[htpb!]
    \centering
    \caption{Cosmological priors used in this work for the $\Lambda$CDM, $w$CDM, and CPL models.
    For all sampled parameters, flat priors $\mathcal{U}[a,b]$ are assumed over the ranges shown,
    except for $\tau_\mathrm{reio}$, which follows a Gaussian prior $\mathcal{N}(\mu,\sigma)$.}
    \renewcommand{\arraystretch}{1.35}
    \resizebox{\columnwidth}{!}{
    \begin{tabular}{l c c c}
        \hline
        {Parameter} & {$\Lambda$}{CDM} & {$w$}{CDM} & {CPL} \\
        \hline\hline
        $\Omega_{\rm b} h^2$              & $\mathcal{U}[0.005,\,0.1]$  & $\mathcal{U}[0.005,\,0.1]$  & $\mathcal{U}[0.005,\,0.1]$  \\
        $\Omega_{\rm c} h^2$              & $\mathcal{U}[0.001,\,0.99]$ & $\mathcal{U}[0.001,\,0.99]$ & $\mathcal{U}[0.001,\,0.99]$ \\
        $100\theta_{\rm MC}$              & $\mathcal{U}[0.5,\,10]$     & $\mathcal{U}[0.5,\,10]$     & $\mathcal{U}[0.5,\,10]$     \\
        $\ln(10^{10} A_{\rm s})$          & $\mathcal{U}[1.61,\,3.91]$  & $\mathcal{U}[1.61,\,3.91]$  & $\mathcal{U}[1.61,\,3.91]$  \\
        $n_{\rm s}$                       & $\mathcal{U}[0.8,\,1.2]$    & $\mathcal{U}[0.8,\,1.2]$    & $\mathcal{U}[0.8,\,1.2]$    \\
        $\tau_{\rm reio}$                 & $\mathcal{N}(0.051,\,0.006)$& $\mathcal{N}(0.051,\,0.006)$& $\mathcal{N}(0.051,\,0.006)$\\
        \hline
        $w_0$                             & $-1$ (fixed)                & $\mathcal{U}[-2,\,0]$       & $\mathcal{U}[-2,\,0]$       \\
        $w_a$                             & $0$ (fixed)                 & $0$ (fixed)                 & $\mathcal{U}[-3,\,2]$       \\
        $\beta_0$                         & $0$ (fixed)                 & $\mathcal{U}[0,\,2]$        & $\mathcal{U}[0,\,2]$        \\
        \hline\hline
    \end{tabular}}
    \label{tab:priors}
\end{table}

Below, we summarize the data sets used in our analysis.

\begin{enumerate}

\item {Cosmic Microwave Background (CMB):}
For the CMB observations, we use the 2018 \textit{Planck} legacy data release, which includes measurements of the temperature and polarization angular power spectra and their cross-correlations~\cite{Planck:2018vyg}. In particular, we adopt the high-$\ell$ \texttt{Plik} likelihood for TT over the multipole interval $30 \leq \ell \leq 2508$, and for TE and EE within $30 \leq \ell \leq 1996$. These measurements are supplemented by the low-$\ell$ temperature likelihood covering $2 \leq \ell \leq 29$, together with the low-multipole polarization likelihood based on \texttt{SimAll} in the same range~\cite{Planck:2019nip}. We further include the CMB lensing likelihood from the \textit{Planck} NPIPE (PR4) data release, reconstructed from the four-point correlation function of the temperature and polarization anisotropies using the quadratic estimator~\cite{Carron:2022eyg}. Throughout this work, this complete set of CMB observations is collectively denoted by {CMB}.

\item {Baryon Acoustic Oscillations (DESI-DR2):}
To probe the expansion history at low and intermediate redshifts, we employ baryon acoustic oscillation (BAO) measurements from the second data release of DESI. The dataset combines BAO information extracted from galaxy and quasar samples~\cite{DESI:2025zgx}, as well as measurements obtained from the Lyman-$\alpha$ forest~\cite{DESI:2025zpo}. The data points, listed in Table IV of Ref.~\cite{DESI:2025zgx}, span the effective redshift interval $0.295 \leq z \leq 2.330$ and are divided into nine redshift bins. The reported observables are the dimensionless quantities $D_{\rm M}/r_d$, $D_{\rm H}/r_d$, and $D_{\rm V}/r_d$, where $D_{\rm M}$ denotes the transverse comoving distance, $D_{\rm H}$ the Hubble distance, $D_{\rm V}$ the volume-averaged distance, and $r_d$ the sound horizon at the baryon drag epoch. The full covariance matrix provided by DESI is incorporated into the likelihood analysis, including the correlations among the BAO observables through the coefficients $r_{\rm V,M/H}$ and $r_{\rm M,H}$. In the following, this dataset is referred to as {DESI-DR2}.

\item {Type Ia Supernovae (SNIa):}
As additional probes of cosmic expansion at late epochs, we consider three recent compilations of Type Ia supernovae. The first is the Pantheon\textsc{Plus} dataset~\cite{Brout:2022vxf}, which comprises 1,701 light-curve measurements corresponding to 1,550 distinct supernovae in the range $0.01 \leq z \leq 2.26$. This sample will henceforth be referred to as {PP}. The second is Union3.0~\cite{Rubin:2023jdq}, containing 2,087 SNIa in the range $0.001 < z < 2.26$, with 1,363 events in common with PP. We designate this dataset as {Union3}. The third is the DES-Dovekie dataset~\cite{DES:2025sig}, a reanalysis of the Dark Energy Survey 5-year sample \cite{DES:2024jxu} based on an updated photometric cross-calibration, comprising 1,623 photometrically classified DES supernovae and $\sim$197 spectroscopically confirmed low-redshift SNe Ia from external surveys (CfA3, CfA4, CSP, and Foundation), for a total of $\sim$1,820 events. While the DES high-redshift component shares no events with PP or Union3, the low-redshift anchor surveys overlap with both. We will refer to this dataset as {DES-DOV}.
\end{enumerate}

\section{Results and discussions}
\label{sec:results}
We now present the constraints obtained for the two cosmology histories introduced in Sec. V, namely the constant-$w$ case ($w$CDM) and the CPL parametrization ($w_0w_a$CDM), each combined with the Nash-gravity scalar sector parametrized by $\beta_0$. The marginalized constraints are summarized in Table~\ref{tab:results} and the contours in Figures~\ref{fig:2D_wcdm}, \ref{fig:2D_CPL} and \ref{fig:results_all}. For each model we also quote $\Delta\chi^2_{\min}$ and $\Delta{\rm AIC}$ relative to $\Lambda$CDM. Since $\beta_0$ is sampled in all cases, the AIC penalty is $2k$ with $k=2$ for $w$CDM and $k=3$ for $w_0w_a$CDM, which is consistent with the entries of Table~\ref{tab:results}.

\begin{table*}[htpb!]
\centering\footnotesize
\caption{Marginalized mean values with $68\%$ confidence intervals for each dataset combination, except for $\beta_0$, for which the $95\%$ CL upper bound is reported. Each block corresponds to a different assumed DE EoS: $w$CDM (first block), $w_aw_0$CDM (second block). The last row of each block reports $\Delta\chi^2_{\rm min}\equiv\chi^2_{\rm min\,(model)}-\chi^2_{\rm min\,(\Lambda CDM)}$ and $\Delta{\rm AIC}\equiv{\rm AIC}_{\rm model}-{\rm AIC}_{\Lambda{\rm CDM}}$. Negative values indicate a preference for the considered model over $\Lambda$CDM.}
\label{tab:results}
\renewcommand{\arraystretch}{1.42}
\resizebox{\textwidth}{!}{
\begin{tabular}{l cccc}
\hline\hline
{Parameter} & {CMB+DESI-DR2} & {CMB+DESI-DR2+PP} & {CMB+DESI-DR2+Union3} & {CMB+DESI-DR2+DES-DOV} \\
\hline
\multicolumn{5}{c}{$w$CDM} \\
\hline
$\Omega_\mathrm{b} h^2$ & $0.02249\pm 0.00013$ & $0.02254\pm 0.00012$ & $0.02253\pm 0.00013$ & $0.02253\pm 0.00012$ \\
$\Omega_\mathrm{c} h^2$ & $0.11842\pm 0.00077$ & $0.11789^{+0.00080}_{-0.00068}$ & $0.11796\pm 0.00078$ & $0.11783^{+0.00070}_{-0.00079}$ \\
$\log(10^{10} A_\mathrm{s})$ & $3.0402\pm 0.0099$ & $3.0423\pm 0.0094$ & $3.0416\pm 0.0097$ & $3.0420\pm 0.0098$ \\
$n_\mathrm{s}$ & $0.9694\pm 0.0035$ & $0.9705\pm 0.0035$ & $0.9703\pm 0.0036$ & $0.9707\pm 0.0035$ \\
$\tau_\mathrm{reio}$ & $0.0535\pm 0.0046$ & $0.0549\pm 0.0042$ & $0.0545\pm 0.0044$ & $0.0547\pm 0.0043$ \\
$100\theta_s$ & $1.04113\pm 0.00028$ & $1.04119\pm 0.00027$ & $1.04119\pm 0.00028$ & $1.04120\pm 0.00027$ \\
$w_0$ & $-1.040\pm 0.035$ & $-0.989\pm 0.022$ & $-0.991^{+0.029}_{-0.026}$ & $-0.984^{+0.023}_{-0.020}$ \\
$\Omega_\mathrm{m}$ & $0.2943\pm 0.0074$ & $0.3047\pm 0.0051$ & $0.3045\pm 0.0061$ & $0.3056\pm 0.0049$ \\
$S_8$ & $0.8106\pm 0.0078$ & $0.8094\pm 0.0079$ & $0.8097\pm 0.0082$ & $0.8091\pm 0.0077$ \\
$H_0$[km/s/Mpc] & $69.37\pm 0.92$ & $68.04\pm 0.57$ & $68.10\pm 0.70$ & $67.93^{+0.50}_{-0.59}$ \\
$\beta_0$ & $< 0.284$ & $< 0.284$ & $< 0.281$ & $< 0.289$ \\
$\Delta\chi^2_\mathrm{min}$ / $\Delta\mathrm{AIC}$ & $-0.43$ / $3.57$ & $-0.04$ / $3.96$ & $-1.31$ / $2.69$ & $-3.50$ / $0.50$ \\
\hline
\multicolumn{5}{c}{$w_aw_0$CDM} \\
\hline
$\Omega_\mathrm{b} h^2$ & $0.02241\pm 0.00013$ & $0.02246\pm 0.00013$ & $0.02243\pm 0.00014$ & $0.02243\pm 0.00013$ \\
$\Omega_\mathrm{c} h^2$ & $0.11951\pm 0.00082$ & $0.11890\pm 0.00081$ & $0.11921\pm 0.00086$ & $0.11914\pm 0.00081$ \\
$\log(10^{10} A_\mathrm{s})$ & $3.0369\pm 0.0094$ & $3.0385\pm 0.0098$ & $3.038\pm 0.010$ & $3.0381\pm 0.0099$ \\
$n_\mathrm{s}$ & $0.9669\pm 0.0034$ & $0.9681\pm 0.0036$ & $0.9675\pm 0.0035$ & $0.9676\pm 0.0034$ \\
$\tau_\mathrm{reio}$ & $0.0515\pm 0.0044$ & $0.0524\pm 0.0046$ & $0.0521^{+0.0050}_{-0.0046}$ & $0.0521\pm 0.0046$ \\
$100\theta_s$ & $1.04096\pm 0.00030$ & $1.04106\pm 0.00028$ & $1.04101\pm 0.00029$ & $1.04101\pm 0.00027$ \\
$w_0$ & $-0.47\pm 0.21$ & $-0.848^{+0.054}_{-0.049}$ & $-0.690\pm 0.088$ & $-0.821^{+0.050}_{-0.059}$ \\
$w_a$ & $-1.55\pm 0.59$ & $-0.54\pm 0.19$ & $-0.97^{+0.30}_{-0.26}$ & $-0.64^{+0.23}_{-0.19}$ \\
$\Omega_\mathrm{m}$ & $0.348\pm 0.021$ & $0.3108\pm 0.0057$ & $0.3256\pm 0.0086$ & $0.3126\pm 0.0051$ \\
$S_8$ & $0.842\pm 0.013$ & $0.8224\pm 0.0090$ & $0.831\pm 0.010$ & $0.8256\pm 0.0088$ \\
$H_0$[km/s/Mpc] & $64.1^{+1.8}_{-2.1}$ & $67.61\pm 0.60$ & $66.13\pm 0.84$ & $67.45\pm 0.52$ \\
$\beta_0$ & $< 0.267$ & $< 0.280$ & $< 0.279$ & $< 0.260$ \\
$\Delta\chi^2_\mathrm{min}$ / $\Delta\mathrm{AIC}$ & $-5.43$ / $0.57$ & $-3.63$ / $2.37$ & $-9.24$ / $-3.24$ & $-9.49$ / $-3.49$ \\
\hline\hline
\end{tabular}}
\end{table*}

\subsection*{\normalfont\textit{Constant-$w$ extrinsic sector}}

For $w$CDM the extrinsic sector reduces to the vacuum Gupta
solution, with $\alpha_0=(1-3w_0)/2$. The data are
fully consistent with a cosmological-constant-like behavior,
$\alpha_0\simeq 2$. 

\begin{figure}[htpb!]
   \centering
   \includegraphics[width=\columnwidth]{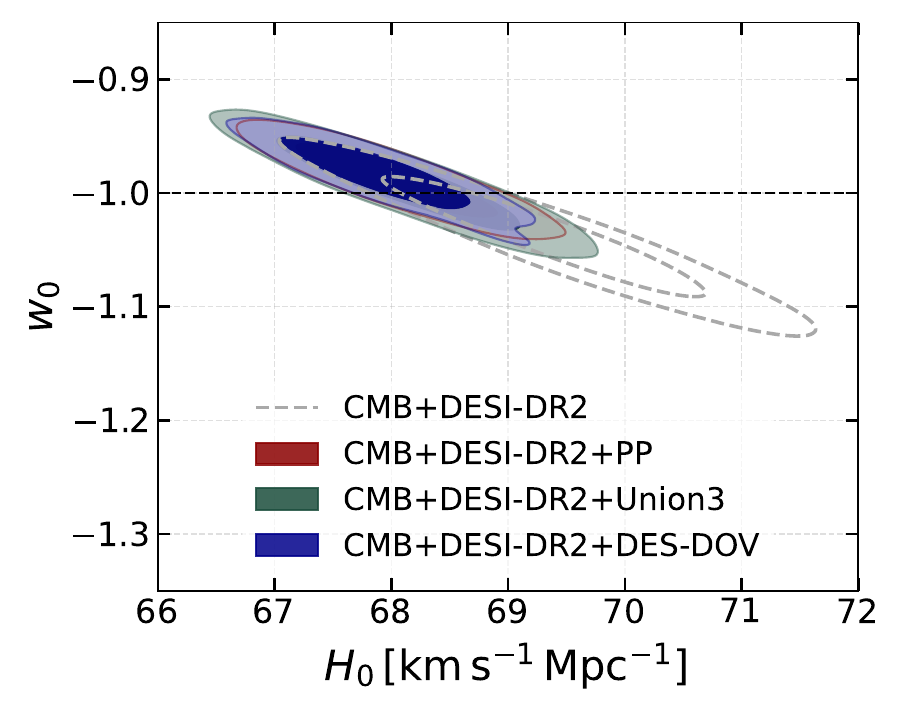} 
   \caption{68\% and 95\% confidence contours in the $w_0$-$H_0$ plane for the $w$CDM model, considering different dataset combinations (see legend).}
   \label{fig:2D_wcdm}
\end{figure}

Figure \ref{fig:2D_wcdm} shows the 2D contours obtained in the $w_0$-$H_0$ plane, with the $\Lambda$CDM model ($w_0=-1$) indicated by a dashed black line. For the CMB+DESI-DR2 data set, we find $w_0=-1.040\pm0.035$, while the inclusion of supernovae shifts the mean to $w_0=-0.989\pm0.022$ (PP), $-0.991^{+0.029}_{-0.026}$ (Union3), and $-0.984^{+0.023}_{-0.020}$ (DES-DOV); all these values lie within approximately $1\sigma$ of $w_0=-1$. The preference for a Hubble constant higher than the value predicted by the $\Lambda$CDM model in the CMB+DESI-DR2 case, with $H_0=69.37\pm0.92$ km\,s$^{-1}$Mpc$^{-1}$, arises from the phantom side of the $w_0$-$H_0$ degeneracy (see Fig.~\ref{fig:2D_wcdm}). This value is reduced to $H_0\simeq 68$ km\,s$^{-1}$Mpc$^{-1}$ when a supernova sample is used to constrain the parameter $\Omega_{\rm m}$ ($\Omega_{\rm m}=0.2943\pm0.0074$ without supernovae, versus $0.305$-$0.306$ with them). The statistical indicators do not favor this model: $\Delta\chi^2_{\min}$ ranges from $-0.04$ to $-3.50$, and $\Delta{\rm AIC}$ is positive in all four cases (from $+0.50$ to $+3.96$). Therefore, the addition of the parameters $w_0$ and $\beta_0$ is not justified by the data.

\subsection*{\normalfont\textit{Dynamical extrinsic sector: CPL}}

\begin{figure}[htpb]
   \centering
   \includegraphics[width=\columnwidth]{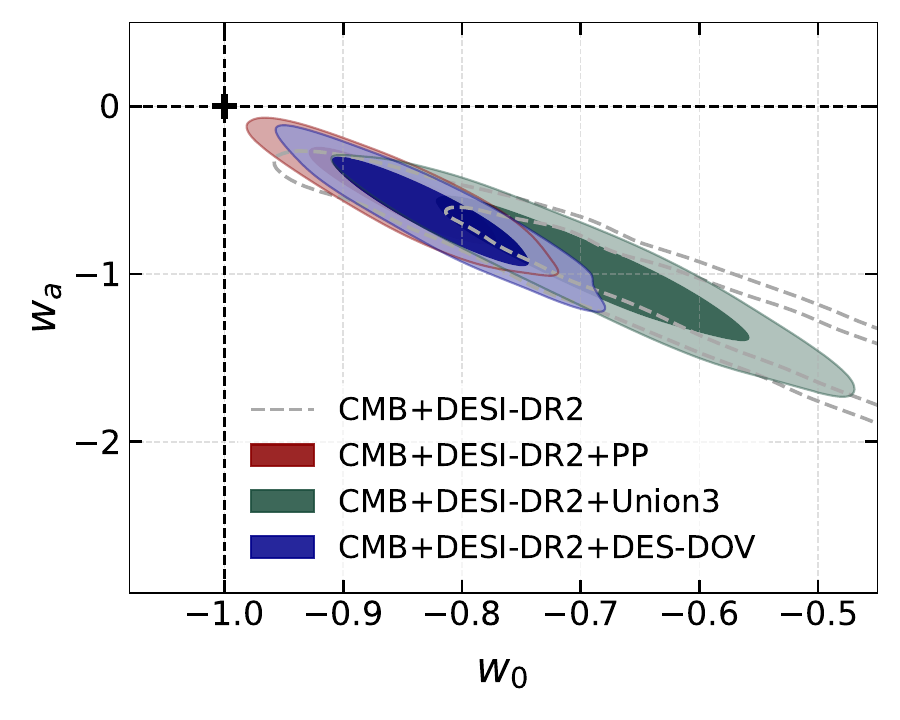} 
   \caption{68\% and 95\% confidence contours in the $w_0$-$w_a$ plane for the CPL model, considering different dataset combinations (see legend).}
   \label{fig:2D_CPL}
\end{figure}

Allowing $\alpha(a)$ to vary through the CPL mapping of Eqs.~\eqref{eq:Xicpl_a}--\eqref{eq:Xicpl_z} changes the picture. In all four dataset combinations the posterior is displaced from the $\Lambda$CDM point $(w_0,w_a)=(-1,0)$ along the usual $w_0$-$w_a$ degeneracy direction (see Fig.~\ref{fig:2D_CPL}). The means are $(w_0,w_a)=(-0.47\pm0.21,\,-1.55\pm0.59)$ for CMB+DESI-DR2, $(-0.848^{+0.054}_{-0.049},\,-0.54\pm0.19)$ with PP, $(-0.690\pm0.088,\,-0.97^{+0.30}_{-0.26})$ with Union3 and $(-0.821^{+0.050}_{-0.059},\,-0.64^{+0.23}_{-0.19})$ with DES-DOV. Taking the marginalized one-dimensional uncertainties at face value, the departures of $w_0$ from $-1$ and of $w_a$ from $0$ correspond to roughly $(2.5, 2.6)\sigma$ for CMB+DESI-DR2 alone and $(3, 2.8)$-$(3.5, 3.4)\sigma$ once supernovae are included. As is well known for CPL fits, these numbers should not be read as a Gaussian significance, given the strongly non-Gaussian, banana-shaped $w_0$-$w_a$ posterior.

\begin{figure*}[htpb!]
    \centering
    \includegraphics[width=0.48\textwidth]{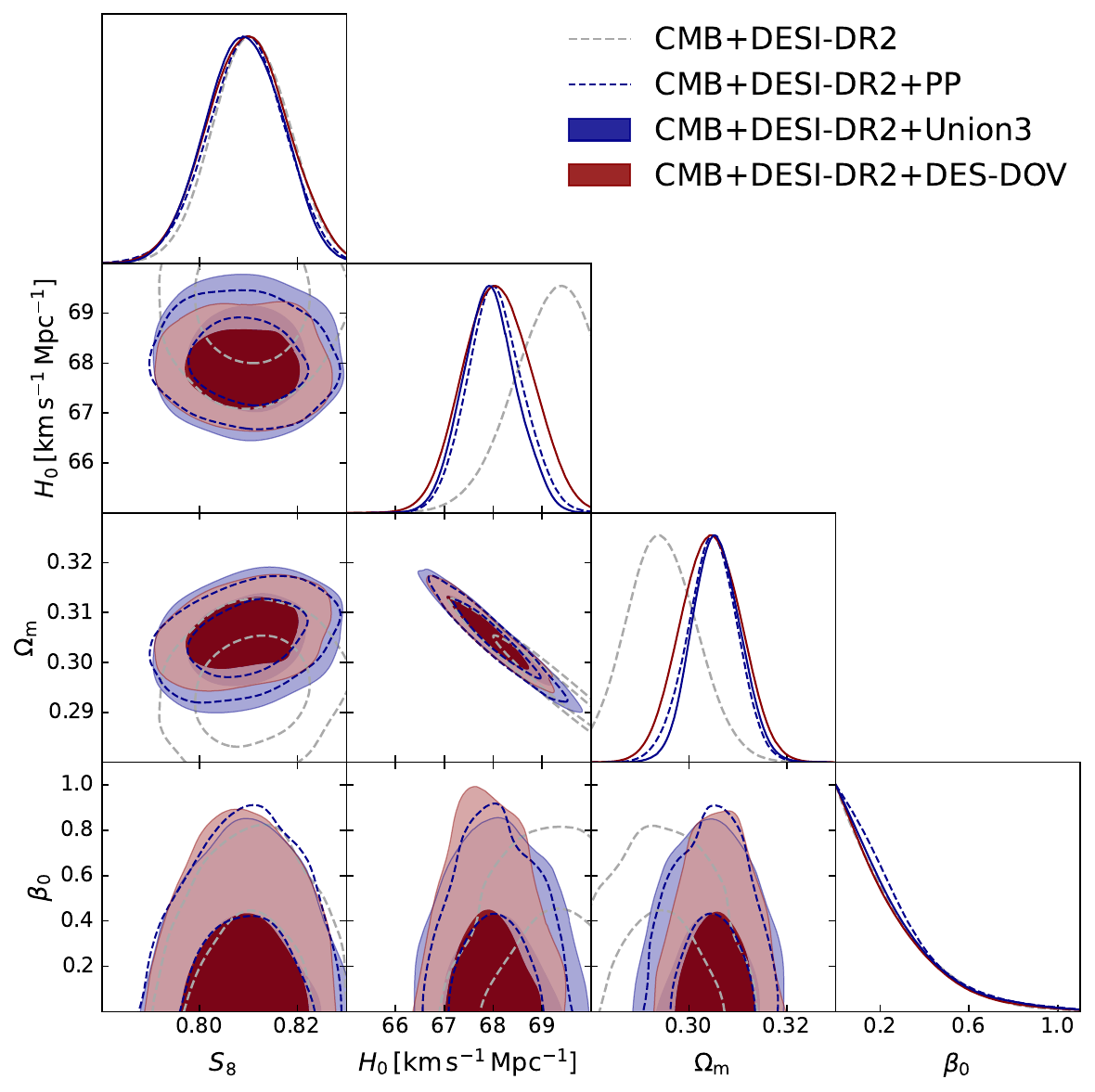} 
    \includegraphics[width=0.48\textwidth]{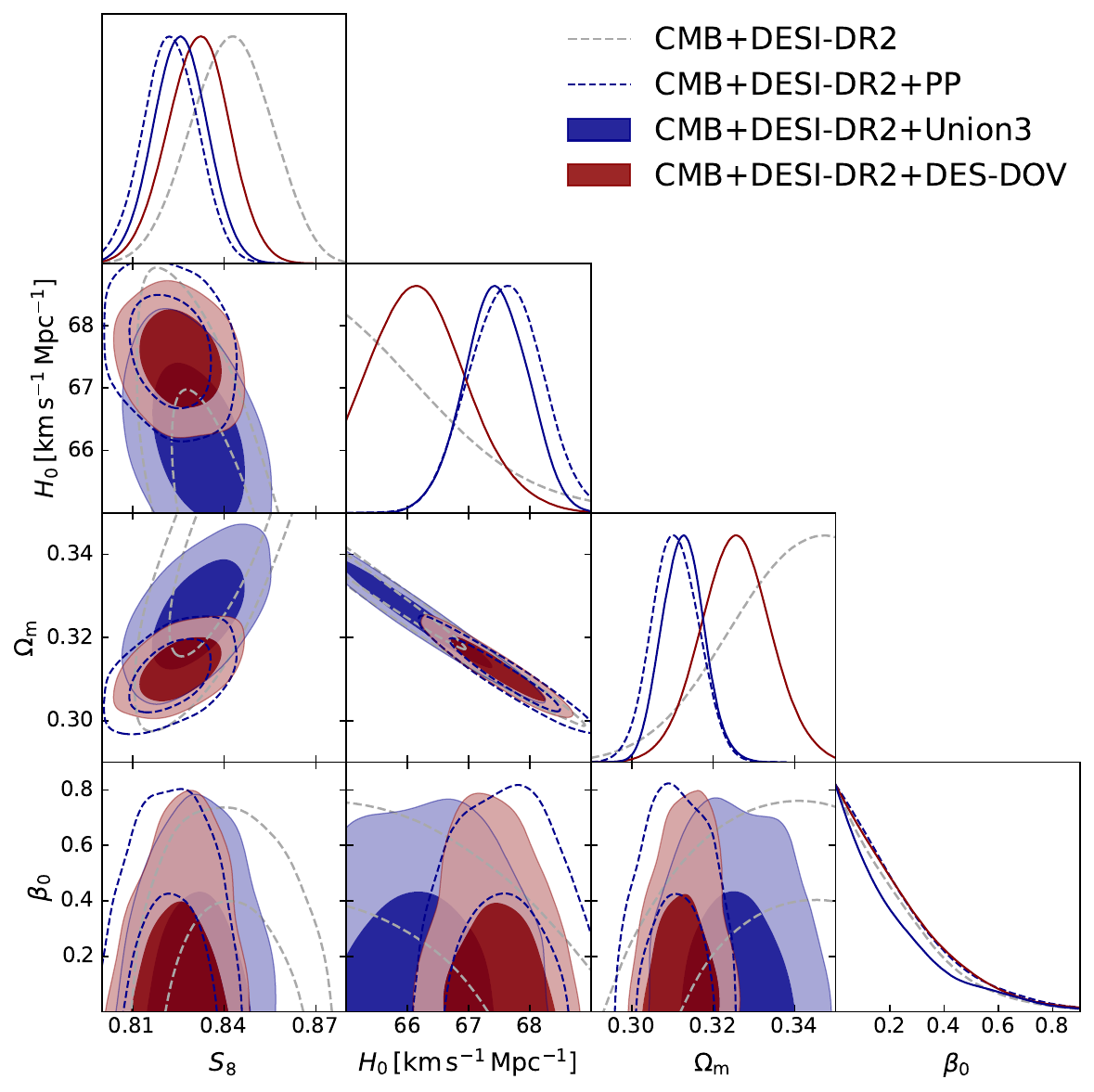} 
    \caption{\textit{Left panel:} Marginalized $68\%$ and $95\%$ confidence contours for the extrinsic $w$CDM scenario. \textit{Right panel:} Same for the extrinsic CPL ($w_0w_a$CDM) scenario. In \textit{both panels}, the four dataset combinations (see legend) are shown in the $S_8$-$H_0$-$\Omega_{\rm m}$-$\beta_0$ parameter space. The $\beta_0$ posterior yields the $95\%$ CL upper bound $\beta_0 \lesssim 0.28$ in both scenarios.}
\label{fig:results_all}
\end{figure*}

Two more comments arise. First, in every case
$w_0+w_a\le -1.39$, so $w_0+w_a<-1/3$ is satisfied automatically by the
data and the extrinsic density is suppressed with respect to matter
at early times while $\mu\to1$ for $a\to0$. Second, the
posterior implies $w(a)=-1$ at $z\simeq 0.4$ - $0.5$
(from $1-a=-(1+w_0)/w_a$: $z\simeq0.52$, $0.39$, $0.47$ and $0.39$ for
the four datasets, respectively), i.e., a phantom divide
crossing from $w<-1$ in the past to $w>-1$ today.
In the Nash language, this corresponds to a bending rate that decreases
across the crossing, $\alpha(a)>2$ at earlier times and
$\alpha_0=(1-3w_0)/2\simeq1.6$ - $1.9$ today (for CMB+DESI-DR2 alone,
$w_0=-0.47$ gives $\alpha_0\simeq1.2$, i.e. a strongly
evolving bending history, driven by the low-$\Omega_{\rm m}$-excluding,
high-$\Omega_{\rm m}=0.348\pm0.021$ solution). This is the first
point in which the extrinsic-curvature description is
genuinely different from the constant-$w$ baseline: no choice of constant $\alpha_0$ can reproduce a crossing.

The statistical improvement is mild but non-negligible. For
CMB+DESI-DR2 alone, $\Delta\chi^2_{\min}=-5.43$ is not enough to
compensate the three extra parameters ($\Delta{\rm AIC}=+0.57$),
and the same holds for PP ($\Delta\chi^2_{\min}=-3.63$,
$\Delta{\rm AIC}=+2.37$). By contrast, the combinations with
Union3 and DES-DOV yield $\Delta\chi^2_{\min}=-9.24$ and
$-9.49$, corresponding to $\Delta{\rm AIC}=-3.24$ and $-3.49$,
which constitute a mild-to-moderate preference over $\Lambda$CDM.

\subsection*{\normalfont\textit{Derived parameters: $H_0$, $\Omega_{\rm m}$ and $S_8$}}

The dynamical models shift the background parameters with
respect to the constant-$w$ case in the direction typical of
DESI-inspired analyses. Our derived parameters are shown for both models in Figure \ref{fig:results_all}. In $w_0w_a$CDM, the
CMB+DESI-DR2 combination prefers a larger matter fraction
$\Omega_{\rm m}=0.348\pm0.021$ and a
lower Hubble constant $H_0=64.1^{+1.8}_{-2.1}$ than in $w$CDM. The
addition of supernovae restores $\Omega_{\rm m}\simeq0.31$ - $0.33$ and
$H_0\simeq66$ - $67.6$ km\,s$^{-1}$Mpc$^{-1}$. Thus, in none of the
analyses does the extrinsic geometry ease the Hubble
tension: values of $H_0$ above $68$ km\,s$^{-1}$Mpc$^{-1}$ appear only in the
$w$CDM case, where they are tied to the phantom side
($w_0=-1.040$) of a model that is not preferred statistically.

Regarding the matter clustering amplitude, the CPL model
yields $S_8=0.822$ - $0.842$ ($w_0w_a$CDM), larger than the $w$CDM values
($0.809$ - $0.811$). Therefore, the better-fitting dynamical
histories move $S_8$ away from the lower values usually
inferred from weak lensing,
and the tension is slightly aggravated rather than alleviated.

\subsection*{\normalfont\textit{Constraints on the Nash amplitude $\beta_0$}}

The scalar sector enters through the single amplitude $\beta_0$
of Eqs.~\eqref{eq:gupta_mode_beta} and \eqref{eq:B_nash_definition}. Its 95\% CL upper bounds are
$\beta_0<0.284$, $0.284$, $0.281$, $0.289$ in $w$CDM and
$<0.267$, $0.280$, $0.279$, $0.260$ in $w_0w_a$CDM, for the
four dataset combinations of Table~\ref{tab:results}.
These limits are remarkably stable as they vary by
less than $\sim10\%$ between models and datasets,
even though the latter differ strongly in their
constraining power on the background. This stability is
expected from the structure of Eq.~\eqref{eq:mu_nash_numerical} with
$\mu=[1+B(a)/k^2]^{-1}$ and $B/(aH)^2\propto\beta_0(3\alpha-1)\Omega_{\rm ext}(a)$,
the modification is active only for $k\lesssim aH$,
i.e., near the horizon, where CMB low-$\ell$ multipoles and the
ISW effect are the only probes available and the constraint
is dominated by cosmic variance. The background data (BAO, SNe)
do not probe this sector, so changing them barely affects $\beta_0$.
It also follows that the bounds are
bounds on a super-horizon effect, and cannot be translated
into constraints on sub-horizon growth, on weak lensing or
on cluster abundances, where $\mu=\Sigma=1$ up to corrections of
order $(aH/k)^2$. Moreover, the posterior for $\beta_0$ is consistent with
$\beta_0=0$ in all cases, and no preference for a nonzero amplitude is
found.

Our main findings are the following:
\begin{itemize}
  \item[-] {Constant-$w$ sector.} A constant-$w$ extrinsic sector is
  consistent with the data but not required: $\Delta{\rm AIC}>0$ in all
  dataset combinations, so $\Lambda$CDM is preferred.

  \item[-] {Dynamical sector.} The CPL histories favor a phantom divide crossing at
  $z\simeq0.4$-$0.5$. The statistical preference over $\Lambda$CDM,
  $\Delta{\rm AIC}\simeq-3$ to $-4$, appears only for the
  Union3 and DES-DOV samples, while CMB+DESI-DR2 alone and
  with PP give no AIC gain. The evidence is therefore moderate and
  supernova-dependent.

  \item[-] {Scalar sector.} The amplitude of the extrinsic scalar mode
  is bounded by $\beta_0\lesssim0.3$ at 95\% CL, with no detection and
  little sensitivity to the background history.
\end{itemize}

\paragraph*{Caveats.}
First, since the framework maps any $w(a)$ onto $\alpha(a)$ through
Eq.~\eqref{eq:alphafromw}, these results test the phenomenology of a dynamical $w(a)$,
as in non-geometric DESI analyses, rather than a specific prediction of
the embedding. A discriminating test would require a first-principles
equation for $S_{\mu\nu}$ or $\alpha(a)$. Second, the stability of the
$\beta_0$ bound indicates that the perturbative sector is
observationally subdominant to the background. A sharper test would
require growth and lensing data (DESI full-shape, KiDS, DES,
HSC) and a scale dependence of $\mu$ and $\Sigma$ extending
into the sub-horizon regime, which Eq.~\eqref{eq:mu_nash_numerical} does not provide. We leave these extensions for future work.

\section{Final remarks}\label{conclusioni}

In this work, we extended the extrinsic-curvature cosmology based on Nash-Greene embeddings beyond the {constant EoS} approximation. {We replaced} the vacuum Ricci-flat Einstein-Gupta closure {with a sourced auxiliary relation, in which the source encodes an unresolved geometric response rather than physical matter.} Under FLRW symmetry, this generalized Gupta sector reduces to a single bending function, {which determines the evolution of the extrinsic density and its effective EoS through the conservation of the extrinsic contribution.} The {constant EoS} model is recovered {when the auxiliary source vanishes and the bending rate becomes constant.} Hence{,} a time-dependent effective EoS arises from the bending dynamics without introducing a new DE fluid or scalar field.

At the perturbative level, we adopted a phenomenological scalar response motivated by the linearized Gupta--Codazzi system, with a vanishing scalar perturbation of the auxiliary source. The prescription yields a scale-dependent metric response and equal growth and lensing modifications when ordinary anisotropic stress is negligible. The background and perturbative responses were thus related through the same bending history and extrinsic density evolution. We then confronted this framework with Planck 2018, DESI DR2 BAO and the PantheonPlus, Union3 and DES-DOV supernova samples, considering both a constant effective EoS and the CPL parametrization of the bending history.

{We found that the constant EoS scenario remained compatible with the observations, although the additional parameters were not statistically favored over the concordance model. Allowing the bending function to evolve according to the CPL parametrization led to a different picture. In particular, the reconstructed histories exhibited a crossing of the phantom divide at redshifts of approximately $0.4 - 0.5$. The corresponding statistical comparison indicated a mild to moderate preference over the concordance model for the combinations including Union3 and DES-DOV. Conversely, the combinations involving PantheonPlus or CMB and DESI alone did not favor the dynamical scenario once the additional parameters were taken into account. Consequently, the statistical preference for an evolving extrinsic contribution depended on the adopted supernova sample.}

{We further examined the implications of the reconstructed histories for the Hubble constant and the matter clustering amplitude. We found that the dynamical scenario did not alleviate the Hubble tension for any of the considered data combinations. Moreover, the inferred clustering amplitude was systematically larger than in the constant EoS case, moving the predictions away from the lower values generally obtained from weak lensing observations. Therefore, although the dynamical extrinsic sector improved the statistical agreement with some combinations of background data, these improvements did not extend to the other cosmological tensions considered in our analysis.}

{Concerning the scalar perturbative sector, we obtained upper bounds on the homogeneous mode amplitude of approximately 0.3 at 95\% confidence, without finding evidence for a nonzero value. These constraints showed only a weak dependence on the assumed background evolution and on the observational datasets. This behavior can be understood from the scale dependence of the modified gravitational response, which becomes relevant mainly near the cosmological horizon and is strongly suppressed on smaller scales. Accordingly, the present bounds primarily constrain the near-horizon contribution and cannot be directly interpreted as constraints on the growth of structures or weak lensing in the subhorizon regime.}

{The present analysis also has two main limitations. First, the correspondence between the bending function and the effective EoS allows us to reproduce a prescribed background evolution, but does not uniquely determine the bending history from the embedding dynamics. The observational constraints therefore test the phenomenology of a dynamical extrinsic contribution without establishing a specific prediction of the underlying geometric theory. Second, the perturbative closure adopted here produces modifications that are predominantly restricted to scales near the cosmological horizon, limiting the possibility of testing the model through the growth of structures.} Future work should therefore (i) derive a dynamical equation for {the auxiliary geometric source or the bending function} from the embedding and (ii) extend the perturbative sector beyond the near-horizon regime so that it {can be tested against} growth and weak-lensing data.

\section*{Acknowledgments}
A.J.S.C.\ acknowledges Conselho Nacional de Desenvolvimento Cient\'{\i}fico e Tecnol\'ogico (CNPq; National Council for Scientific and Technological Development) for partial financial support (Grant No.~305881/2022-1) and Funda\c{c}\~ao da Universidade Federal do Paran\'a (FUNPAR; Paran\'a Federal University Foundation) through public notice 04/2023-Pesquisa/PRPPG/UFPR for partial financial support (Process No.~23075.019406/2023-92), as well as the financial support of the NAPI ``Fen\^omenos Extremos do Universo'' of Funda\c{c}\~ao de Apoio \`a Ci\^encia, Tecnologia e Inova\c{c}\~ao do Paran\'a (NAPI F\'ISICA-FASE~2), under protocol No.~22.687.035-0. R.C.N. thanks the financial support from the CNPq under the project No. 304306/2022-3, and the Fundação de Amparo à Pesquisa do Estado do RS (FAPERGS, Research Support Foundation of the State of RS) for partial financial support under the project No. 23/2551-0000848-3. E.S. received support from the CAPES (Coordination for the Improvement of Higher Education Personnel) scholarship. O.L. thanks Roberto Della Ceca and the Brera National Institute of Astrophysics for support and acknowledges the Istituto Nazionale di
Fisica Nucleare (INFN) through the Iniziativa Specifica
QGSKY.  SK acknowledges support of the Startup Research Grant from Plaksha University (File No.\ OOR/PU-SRG/2023-24/08).

\end{document}